\newcommand*{\ATLASLATEXPATH}{latex/}
  
\documentclass[cernpreprint,UKenglish,texlive=2016,txfonts]{\ATLASLATEXPATH atlasdoc}

\pdfoutput=1

\usepackage{\ATLASLATEXPATH atlaspackage}
\usepackage{lscape}
\usepackage{verbatim}
\usepackage{\ATLASLATEXPATH atlasbiblatex}

\usepackage{\ATLASLATEXPATH atlasphysics}
\usepackage{\ATLASLATEXPATH atlasprocess}

\usepackage{ZplusJetsPaper-def}
\usepackage{color}
\usepackage{atlas_tables-defs}
\usepackage{siunitx}
 
\graphicspath{{logos/}{figures/}}

\AtlasTitle{Measurements of the production cross section of a \Zboson boson in association
with jets in \textit{pp} collisions at $\sqrt{s} = 13$~TeV with the ATLAS detector}

\author{The ATLAS Collaboration}

\AtlasRefCode{STDM-2016-01}

\PreprintIdNumber{CERN-EP-2016-297}

\AtlasJournalRef{Eur. Phys. J. C77 (2017) 361} 
\AtlasDOI{10.1140/epjc/s10052-017-4900-z}

\AtlasAbstract{
Measurements of the production cross section of a \Zboson boson in association with jets in proton--proton collisions at $\sqrt{s} = 13$~TeV are presented, using data corresponding to an integrated luminosity of $ 3.16$~fb$^{-1}$ collected by the ATLAS experiment at the CERN Large Hadron Collider in 2015. Inclusive and differential cross sections are measured for events containing a \Zboson boson decaying to electrons or muons and produced in association with up to seven jets with $\pT > 30$~GeV and $|y| <2.5$. Predictions from different Monte Carlo generators based on leading-order and next-to-leading-order matrix elements for up to two additional partons interfaced with parton shower and fixed-order predictions at next-to-leading order and next-to-next-to-leading order are compared with the measured cross sections. Good agreement within the uncertainties is observed for most of the modelled quantities, in particular with the generators which use next-to-leading-order matrix elements and the more recent next-to-next-to-leading-order fixed-order predictions.\\

}

\hypersetup{pdftitle={ATLAS draft},pdfauthor={The ATLAS Collaboration}}

\newcommand{\ZNj}{\mbox{$Z (\rightarrow \ell^{+}\ell^{-})+\ge N_{\mathrm{{jets}}}$}}

\begin{document}

\maketitle

\tableofcontents
\clearpage

\section{Introduction}\label{sec:intro}
 
The measurement of the production of a \Zboson boson\footnote{Throughout this paper, \Zg-boson production is denoted simply by \Zboson-boson production.}  in association with jets, \Zj,  constitutes a powerful test of perturbative quantum chromodynamics (QCD)~\cite{QCD1,QCD2}. The large production cross section and easily identifiable decays of the \Zboson boson to charged leptonic final states offer clean experimental signatures which can be precisely measured.  Such processes also constitute a non-negligible  background for studies of the Higgs boson and in searches for new phenomena; typically in these studies,  the multiplicity and kinematics of the jets are exploited  to achieve a separation of the signal of interest from the Standard Model (SM)  \Zj process. These quantities are often measured in control regions and subsequently extrapolated to the signal region with the use of Monte Carlo  (MC) generators, which are themselves subject to systematic uncertainty and must be tuned and validated using data. Predictions from the most recent generators combine next-to-leading-order (NLO) multi-leg matrix elements with a parton shower (PS) and a hadronisation model. Fixed-order parton-level predictions for \Zj production at next-to-next-to-leading order (NNLO) are also available~\cite{Boughezal:2015ded,Boughezal:2016isb,Ridder:2015dxa,Ridder:2016nkl}.

The \Zj production differential cross section was previously measured by the ATLAS~\cite{PERF-2007-01}, CMS~\cite{Chatrchyan:2008aa}, and LHCb~\cite{Alves:2008zz} collaborations at the CERN Large Hadron Collider (LHC)~\cite{1748-0221-3-08-S08001} at centre-of-mass energies of $\sqrt{s}=7$~TeV~\cite{Zjets2013,VjetsCms1,VjetsCms2,VjetsCms3,Aaij:2013nxa} and $8$~TeV~\cite{Khachatryan:2015ira,Khachatryan:2016crw,AbellanBeteta:2016ugk}, and by the CDF and D0 collaborations at the Tevatron collider at $\sqrt s = 1.96$~TeV~\cite{Aaltonen:2014vma,Abazov:2009pp}. In this paper, proton--proton ($pp$) collision data corresponding to an integrated luminosity of 3.16~fb$^{-1}$, collected at $\sqrt{s} = 13$~TeV with the ATLAS detector during 2015, are used for measurements of the \Zboson-boson production cross section in association with up to seven jets within a fiducial region defined by the detector acceptance. The \Zboson boson is identified using its decays to electron or muon pairs  (\Ztoee, \Ztomm). Cross sections are measured separately for these two channels, and for their combination, as a function of the inclusive and exclusive jet multiplicity $N_\mathrm{jets}$ and the ratio of successive inclusive jet multiplicities $(N_\mathrm{jets}+1)/N_\mathrm{jets}$, the transverse momentum  of the leading jet \pTj for several jet multiplicities, the jet rapidity \yj, the azimuthal separation between the two leading jets \Dphijj, the two leading jet invariant mass \mjj, and the scalar sum \HT of the transverse momenta of all selected leptons and jets. 

The paper is organised as follows. Section~\ref{sec:detector} contains a brief description of the ATLAS detector. The data and simulated samples as well as the \Zj predictions used in the analysis are described in Section~\ref{sec:dataMC}. The event selection and its associated uncertainties are presented in Section~\ref{sec:selection}, while the methods employed to estimate the backgrounds are shown in Section~\ref{sec:backgrounds}. Comparisons between data and Monte Carlo predictions for reconstructed distributions are found in Section~\ref{sec:kinematics}, while the unfolding procedure is described in Section~\ref{sec:unfolding}. Section~\ref{sec:results} presents the analysis results, the comparisons to predictions, and a discussion of their interpretation. Conclusions are provided in Section~\ref{sec:conclusion}.

\section{The ATLAS detector}
\label{sec:detector}

\newcommand{\AtlasCoordFootnote}{%
ATLAS uses a right-handed coordinate system with its origin at the nominal interaction point (IP)
in the centre of the detector and the $z$-axis along the beam pipe.
The $x$-axis points from the IP to the centre of the LHC ring,
and the $y$-axis points upwards.
Cylindrical coordinates $(r,\phi)$ are used in the transverse plane, 
$\phi$ being the azimuthal angle around the $z$-axis.
The pseudorapidity is defined in terms of the polar angle $\theta$ as $\eta = -\ln \tan(\theta/2)$.
Angular distance is measured in units of $\Delta R \equiv \sqrt{(\Delta\eta)^{2} + (\Delta\phi)^{2}}$.
When dealing with massive jets and particles, the rapidity $y = \frac{1}{2} \ln \left( \frac{E+p_z}{E-p_z} \right)$
is used, where $E$ is the jet/particle energy and $p_z$ is the $z$-component of
the jet/particle momentum. }

The ATLAS experiment at the LHC is a multi-purpose particle detector with a forward-backward symmetric cylindrical geometry and nearly $4\pi$ coverage in  solid angle.\footnote{\AtlasCoordFootnote} It consists of an inner tracking detector, electromagnetic and hadronic calorimeters, and a muon spectrometer. 
The inner tracker is surrounded by a thin superconducting solenoid magnet and provides precision tracking of charged particles and momentum measurements in the pseudorapidity range $|\eta| < 2.5$.  
This region is matched to a high-granularity 
electromagnetic (EM) sampling calorimeter covering the pseudorapidity range $|\eta| < 3.2$, and a coarser granularity calorimeter up to $|\eta| = 4.9$. 
The hadronic calorimeter system covers the entire pseudorapidity range up to  $|\eta| = 4.9$.
The muon spectrometer consists of three large superconducting toroids each containing eight coils, a system of trigger chambers, and precision tracking
chambers, which provide trigger and tracking capabilities in the range $|\eta| < 2.4$ and $|\eta| < 2.7$, respectively.
A two-level trigger system~\cite{Aaboud:2016leb} is used to select events. The first-level
trigger is implemented in hardware and uses a subset of the detector
information. This is followed by the software-based high-level trigger system,
which runs offline reconstruction, reducing the event rate to approximately \SI{1}{kHz}.

\section{Data set, simulated event samples, and predictions}
\label{sec:dataMC}

\subsection{Data set}

The data used in this analysis were collected by the ATLAS detector during August  to November 2015. During this period, the LHC circulated \SI{6.5}{\TeV} proton beams with a \SI{25}{ns} bunch spacing.  The peak delivered instantaneous luminosity was $L=\SI{5e33}{\per \cm \squared \per \s}$ and the mean number of $pp$ interactions per bunch crossing (hard scattering and pile-up events) was $\langle \mu \rangle = 13$.  The data set used in this measurement corresponds to a total integrated luminosity of 3.16~fb$^{-1}$.

\subsection{Simulated event samples}

Monte Carlo simulations, normalised to higher-order calculations,  are used to estimate most of the contributions from background events, to unfold the data to the particle level, and to compare with the unfolded data distributions. All  samples are processed with a {\textsc{Geant}4}-based simulation~\cite{Agostinelli:2002hh} of the ATLAS detector~\cite{SOFT-2010-01}. An overview of all signal and background processes considered and of the generators used for the simulation is given in Table~\ref{tab:MC}.  Total production cross sections for the samples, their respective uncertainties (mainly coming from parton distribution function (PDF) and factorisation and renormalisation scale variations), and references to higher-order QCD corrections, where available, are also listed in Table~\ref{tab:MC}.

\begin{table}[t]
\begin{scriptsize}
\begin{center} 
\begin{tabular}{ |l|l|rl c|r|}
\hline
\hline
Process & Generator& $(\sigma \cdot \mathrm{BR})$ & Normalisation & Reference & Theory\\ 
        &          & [pb]                    & order &                   & uncert.\\
\hline
\hline
 \Ztollj ($\ell = e,\mu; m_{\ell\ell}>40$~GeV) &     \SHERPAtwotwo & $2106$  &NNLO &\cite{Anastasiou:2003ds,FEWZ2,FEWZ3,FEWZ4} & 5\% \\
 \Ztollj ($\ell = e,\mu,\tau; m_{\ell\ell}>40$~GeV) &     \MGaMCPyeight          & $2103$  &NNLO & \cite{Anastasiou:2003ds,FEWZ2,FEWZ3,FEWZ4} & 5\%  \\
 \Wln\ \;\;($\ell = e,\mu$)   	            &   \MGaMCPyeight    & $20080$  &NNLO &  \cite{Anastasiou:2003ds,FEWZ2,FEWZ3,FEWZ4} & 5\%   \\
$t\overline{t}$ \;\; ($m_{t}=172.5$~GeV) & & & & & \\           		
\;\;  \textsc{Perugia2012(radHi/radLo)}  &\POWHEGPysix   & $831$ &NNLO+NNLL &\cite{Czakon:2011xx} & 6\%\\
\;\; \textsc{UE-EE-5}                                        &\MGaMCHpp  & $831$ & NNLO+NNLL & \cite{Czakon:2011xx} & 6\%\\
Single top quark ($Wt$)                                   &\POWHEGPysix   & $72$         & NLO+NNLL &\cite{Kidonakis2010u} & 6\% \\
Single top quark ($t$-channel)                             &\POWHEGPysix   & $136$   & NLO+NNLL &\cite{Kidonakis2011} & 6\%\\
Single top anti-quark ($t$-channel)                             &\POWHEGPysix   & $81$   & NLO+NNLL &\cite{Kidonakis2011} & 6\%\\
Dibosons                                                      & \SHERPAtwoone& $97$ & NLO & \cite{Gleisberg:2008ta} & 6\% \\
\hline
\hline
\end{tabular}
\caption{Signal and background Monte Carlo samples and the generators used in the simulation. 
Each sample is normalised to the appropriate production cross section $\sigma$ and  multiplied by the relevant branching ratios (${\mathrm{BR}}$) per lepton flavour for this sample, as shown in the third column.  
For \Wboson-boson and top-quark production, contributions from higher-order QCD corrections were calculated  following the references given in the fifth column for the stated order. Similarly, for \Zboson-boson production, higher-order QCD corrections were evaluated in the dilepton invariant mass range $66<m_{\ell\ell}<116$~GeV following the references given in the fifth column for the stated order, and extrapolation scaling factors were applied to match mass ranges used by each simulation as given in the first column.  The theory uncertainties as given in the final column correspond to PDF and scale variations. The diboson samples include on-shell and off-shell $WW$, $WZ$ and $ZZ$ production. Recently, NNLO QCD predictions have been made available for the diboson processes~\cite{wznnlo,ggwwnnlo}. However, these higher-order corrections have a negligible impact on this analysis. }
\label{tab:MC}
\end{center}
\end{scriptsize}
\end{table}

Signal events (i.e. containing a \Zboson boson with associated jets) are simulated using the \SHERPA~v2.2.1~\cite{Gleisberg:2008ta} generator, denoted by \SHERPAtwotwo. Matrix elements (ME) are calculated for up to two additional partons at NLO and up to four partons at leading order (LO) using the  Comix~\cite{Gleisberg:2008fv} and OpenLoops~\cite{Cascioli:2011va} matrix element generators. They are merged with the \SHERPA parton shower~\cite{Schumann:2007mg} (with a matching scale of 20~GeV) using the \textsc{ME+PS@NLO} prescription~\cite{Hoeche:2012yf}. A five-flavour scheme is used for these predictions. The \NNPDFthree PDF set~\cite{Ball:2014uwa} is used in conjunction with a dedicated set of parton-shower-generator parameters (tune) developed by the \SHERPA authors.
This sample is used for the nominal unfolding of the data distributions, to compare to the cross-section measurements, and to estimate the systematic uncertainties.

A simulated sample of \Zj production is also produced with the \MADGRAPHaMC (denoted by \MGaMC)~v2.2.2 generator~\cite{Alwall:2014hca}, using matrix elements including up to four partons at leading order and employing the \NNPDFthree PDF set,  interfaced to \PYTHIA~v8.186~\cite{Sjostrand:2007gs} to model the parton shower, using the CKKWL merging scheme~\cite{Lonnblad:2001iq} (with a matching scale of 30~GeV).  A five-flavour scheme is used. The \textsc{A14}~\cite{ATL-PHYS-PUB-2014-021} parton-shower tune is used together with the \NNPDFtwothree PDF set~\cite{Ball:2012cx}. The sample is denoted by \MGaMCPyCKKWL and is used to provide cross-checks of the systematic uncertainty in the unfolding and to model the small $\Zboson \to \tau \tau$ background. In addition, a \MGaMC sample with matrix elements for up to two jets and with parton showers beyond this, employing the \NNPDFthree PDF set and interfaced to \PYTHIA~v8.186 to model the parton shower, is generated using the FxFx merging scheme~\cite{VjetFxFx}  (with a matching scale of 25~GeV~\cite{VjetFxFxold}) and is denoted by \MGaMCPyFxFx. This sample also uses a five-flavour scheme and the \textsc{A14} parton-shower tune  with the \NNPDFtwothree PDF set. Both \MGaMC samples are used for comparison with  the unfolded cross-section measurements.

The measured cross sections are also compared to predictions from the leading-order matrix element generator \ALPGEN~v2.14~\cite{Mangano:2002ea} interfaced to 
\PYTHIA~v6.426~\cite{Sjostrand:2006za} to model the parton shower, denoted by \ALPGENPysix, using the \textsc{Perugia2011C}~\cite{Skands:2010ak} parton-shower tune and 
the  \textsc{CTEQ6L1} PDF set~\cite{Pumplin:2002vw}. A four-flavour scheme is used. Up to five additional partons are modelled by the matrix elements merged with the MLM 
prescription~\cite{Mangano:2002ea} (with a matching scale of 20~GeV).  The matrix elements for the production of $Z+b\bar{b}$ and $Z+c\bar{c}$ events are
 explicitly included and a heavy-flavour  overlap procedure is used to remove the double counting of heavy quarks from gluon splitting in the parton shower.

The \Zboson-boson samples are normalised to the NNLO prediction calculated with the {\textsc Fewz}~3.1 program~\cite{Anastasiou:2003ds,FEWZ2,FEWZ3,FEWZ4} with \textsc{CT10nnlo} PDFs~\cite{Lai:2010vv}. 

Contributions from the top-quark, single-boson, and diboson components of the background (described in Section~\ref{sec:backgrounds}) are estimated from the following Monte Carlo samples.
Samples of top-quark  pair and single top-quark production are generated at NLO with the \POWHEG-Box generator~\cite{Nason:2004rx,Frixione:2007vw,Alioli:2010xd} (versions~v2 (r3026) for top-quark pairs and~v1 for single top quarks (r2556 and r2819 for $t$-and $Wt$-channels, respectively)) and \PYTHIA~v6.428 (\textsc{Perugia2012} tune~\cite{Skands:2010ak}). Samples with enhanced or suppressed additional radiation are generated with the \textsc{Perugia2012radHi}/\textsc{Lo} tunes~\cite{Skands:2010ak}. An alternative top-quark pair sample is produced using the \MGaMC generator interfaced with \HERWIGpp~v2.7.1~\cite{Alwall:2014hca,herwigpp}, using the \textsc{UE-EE-5} tune~\cite{Gieseke:2012ft}. The samples are normalised to the cross section calculated at NNLO+NNLL (next-to-next-to-leading log) with the Top++2.0 program~\cite{Czakon:2011xx}.

The \Wboson-boson backgrounds are modelled using the \MGaMCPyCKKWL~v2.2.2 generator, interfaced to \PYTHIA~v8.186 and are normalised to the NNLO values given in Table~\ref{tab:MC}.
Diboson processes with fully leptonic and semileptonic decays are simulated~\cite{MultiBosonPub} using the \SHERPA~v2.1.1 generator with the \textsc{CT10nlo} PDF set. The matrix elements contain the doubly resonant $WW$, $WZ$ and $ZZ$ processes, and all other diagrams with four electroweak vertices. They are calculated for one or zero additional partons at NLO and up to three additional partons at LO and merged with the \SHERPA parton shower using the \textsc{ME+PS@NLO} prescription. 

Events involving semileptonic decays of heavy quarks, hadrons misidentified as leptons, and, in the case of the electron channel, electrons from photon conversions are referred to collectively as ``multijet events''.  The multijet background was estimated using data-driven techniques, as described in Section~\ref{sec:backgrounds}.

Multiple overlaid $pp$ collisions are simulated with the soft QCD processes of \PYTHIA~v.8.186 using the \textsc{A2} tune~\cite{ATLAS-PHYS-PUB-2012-003} and the \textsc{MSTW2008LO} PDF set~\cite{Martin:2009iq}. All Monte Carlo samples are reweighted so that the pile-up distribution matches that observed in the data.  

\subsection{Fixed-order predictions}
\label{sec:dataMCFO}
    
In addition to these Monte Carlo samples, parton-level fixed-order predictions at NLO are calculated by the \BLACKHATSHERPA collaboration for the production of \Zboson bosons with  up to  four partons~\cite{bhz3jets,bhz4jets}. 
The \BLACKHATSHERPA predictions use the \textsc{CT14} PDF set~\cite{Dulat:2015mca} including variations of its eigenvectors at the 68\% confidence level, rescaled from 90\% confidence level using a factor of $1/1.645$. 
The nominal predictions use a factorisation and renormalisation scale of $\HT/2$ with uncertainties derived from the envelope of a common variation of the scales by factors of $0.5, 1/\sqrt{2}, \sqrt{2},$ and 2.
The effects of PDF and scale uncertainties range from 1\% to 4\% and from 0.1\% to 10\%, respectively, for the cross sections of \Zboson-boson production in association with at least one to four partons, and are included in the predictions which are provided by the  \BLACKHATSHERPA authors for the fiducial phase space of this analysis. 
Since these predictions are defined before the decay leptons emit photons via final-state radiation (Born-level FSR), corrections to the dressed level (where all photons found within a cone of size $\Delta R = 0.1$ of the lepton from the decay of the \Zboson boson are included) are derived from \MGaMCPyCKKWL, separately for each kinematic observable used to measure cross sections,  with associated systematic uncertainties obtained by comparing to the \ALPGENPysix generator.
This correction is needed in order to match the prediction to the lepton definition used in the measurements. The average size of these corrections is approximately $-2$\%. 
To bring the prediction from parton to particle level, corrections for the non-perturbative effects of hadronisation and the underlying event are also calculated separately for each observable  using the \SHERPA~v2.2 generator by turning on and off in the simulation both the fragmentation and the interactions between the proton remnants. The net size of the corrections is up to approximately 10\% at small values of \pTj and vanishes for large values of \pTj. An uncertainty of approximately 2\% for this correction is included in  the total systematic uncertainty of the prediction. 

Calculations of cross sections at NNLO QCD have recently become available~\cite{Boughezal:2015ded,Boughezal:2016isb,Ridder:2015dxa,Ridder:2016nkl}. In this paper, the results are compared to the calculation, denoted by \Njetti\cite{Boughezal:2015ded,Boughezal:2016isb}, which uses a new subtraction technique based on $N$-{\em jettiness}~\cite{Boughezal:2015dva} 
 and relies on the theoretical formalism provided in soft-collinear effective theory. The predictions, which are provided by the authors of this calculation for the fiducial phase space of this analysis, use a factorisation and renormalisation scale of $\sqrt{m_{\ell\ell}^2 + \sum \left(p_{\text{T}}^{\text{jet}}\right)^2}$ (where $m_{\ell\ell}$ is the invariant mass of the dilepton system) and the \textsc{CT14} PDF set.  The QCD renormalisation and factorisation scales were jointly varied by a common factor of two, and are included in the uncertainties. Non-perturbative and FSR corrections and their associated uncertainties as discussed above are also included in the predictions.

\section{Event selection}
\label{sec:selection}

Electron- and muon-candidate events are selected using triggers which require at least one electron or muon with transverse momentum thresholds of $\pT=\SI{24}{\GeV}$ or \SI{20}{\GeV}, respectively, with some isolation requirements for the muon trigger. To recover possible efficiency losses at high momenta, additional electron and muon triggers which do not make any isolation requirements are included with thresholds of $\pT \ge \SI{60}{\GeV}$ and $\pT = \SI{50}{\GeV}$, respectively.  Candidate events are required to have a primary vertex, defined as the vertex with the highest sum of track $\pT^2$, with at least two associated tracks with $\pT > \SI{400}{\MeV}$.

Electron candidates are required to have $\pT>\SI{25}{\GeV}$ and to pass ``medium'' likelihood-based identification requirements~\cite{elecperf,ATL-PHYS-PUB-2015-041} optimised for the 2015 operating conditions, within the fiducial region of $|\eta|<2.47$, excluding candidates in the transition region between the barrel and endcap electromagnetic calorimeters, $1.37<|\eta|<1.52$. Muons are reconstructed for $|\eta|<2.4$ with $\pT>\SI{25}{\GeV}$ and must pass ``medium'' identification requirements~\cite{Aad:2016jkr} also optimised for the 2015 operating conditions. At least one of the lepton candidates is required to match the lepton that triggered the event. The electrons and muons must also satisfy \pT-dependent cone-based isolation requirements, using both tracking detector and calorimeter information (described in Refs.~\cite{Aad:2015yja} and ~\cite{Aad:2014lwa}, respectively). The isolation requirements are tuned so that the lepton isolation efficiency is at least 90\% for $\pT>\SI{25}{\GeV}$, increasing to 99\% at 60~GeV.
Both the electron and muon tracks are required to be associated with the primary vertex, using constraints on the transverse impact parameter significance $|d_0|/\Delta d_0$, where $d_0$ is the transverse impact parameter and $\Delta d_0$ is its uncertainty, and on the longitudinal impact parameter $z_0$ corrected for the reconstructed position of the primary vertex. The transverse impact parameter significance is required to be less than five for electrons and three for muons, while the absolute value of the corrected $z_0$ multiplied by the sine of the track polar angle is required to be less than 0.5~mm.

Jets of hadrons are reconstructed with the anti-$k_t$ algorithm~\cite{Cacciari:2008} with radius parameter $R=0.4$ using  topological clusters of energy deposited in the calorimeters~\cite{Aad:2016upy}. Jets are calibrated using a simulation-based calibration scheme, followed by in situ corrections to account for differences between simulation and data~\cite{ATLAS-PHYS-PUB-2015-015}. In order to reduce the effects of pile-up contributions, jets with pseudorapidity  $|\eta|<2.4$ and  $\pT<\SI{60}{\GeV}$ are required to have a significant fraction of their tracks with an origin compatible with the primary vertex, as defined by the jet vertex tagger algorithm~\cite{ATLAS-CONF-2014-018}. In addition, the expected average energy contribution from pile-up clusters is subtracted according to the $\eta$--$\phi$ catchment area of the jet~\cite{Aad:2015ina}. Jets used in the analysis are required to have  \pT greater than $30$~GeV and  rapidity $|y|<2.5$. 

The overlap between leptons and jets is removed in a two-step process. The first step removes jets closer than $\Delta R  = 0.2$ to a selected electron, and jets closer than $\Delta R  = 0.2$ to a selected muon, if they are likely to be reconstructed from  photons radiated by the muon. 
In a second step,  electrons and muons are discarded if they are located closer than $\Delta R = 0.4$ to a remaining selected jet.
This requirement effectively removes events with leptons and jets which are not reliably simulated in the Monte Carlo simulation.

Events containing a \Zboson-boson candidate are selected by requiring exactly two leptons of the same flavour but of opposite charge with dilepton invariant mass in the range  $71 < m_{\ell\ell} < \SI{111}{GeV}$. The expected and observed numbers  of \Zboson-boson candidates selected for each inclusive jet multiplicity, for $N_{\mathrm{jets}} \ge 0-7$, are summarised in Table~\ref{tab:zjets_sigbak}, separately for the \Ztoee and the \Ztomm channels. The background evaluation that appears in this table is discussed in Section~\ref{sec:backgrounds}. After all requirements,  248,816   and  311,183  \Zonej events are selected in the electron and muon channels, respectively.

\begin{table}[t]
\centering
\begin{footnotesize}
\begin{tabular}{|l|r|r|r|r|r|r|r|r|}
\hline \hline
& \multicolumn{8}{c|}{Electron channel} \\ \hline
& $\texttt{+}\geq\mathrm{0~jets} $ & $\texttt{+}\geq\mathrm{1~jets}$ & $\texttt{+}\geq\mathrm{2~jets}$ & $\texttt{+}\geq\mathrm{3~jets}$ & $\texttt{+}\geq\mathrm{4~jets}$ & $\texttt{+}\geq\mathrm{5~jets}$ & $\texttt{+}\geq\mathrm{6~jets}$ & $\texttt{+}\geq\mathrm{7~jets}$ \\
\hline
\Ztoee\ [\%]               & 99.3 & 97.6 & 93.9 & 90.3 & 87.3 & 85.2 & 83.3 & 81.2\\
Top quark   [\%]           & 0.2  & 1.2 & 3.8  & 6.5  & 8.6  & 9.7 & 10.5 & 11.6\\
Diboson   [\%]             & 0.2  & 0.8  & 1.6  & 2.4  & 3.4  & 4.4  & 5.5  & 6.6\\
$Z\to\tau^{+}\tau^{-}$ [\%] & <0.1 & <0.1 & <0.1 & <0.1 & <0.1 & <0.1 & <0.1 & <0.1\\
$W \to e\nu$  [\%]         & <0.1 & <0.1 & <0.1 & <0.1 & <0.1 & <0.1 & <0.1 & <0.1\\ 
Multijet      [\%]         & 0.2  & 0.4  & 0.6  & 0.7  & 0.7  & 0.7  & 0.7  & 0.7\\
\hline  

Expected               & 1,327,900 & 239,500 & 57,310 & 14,080 & 3637 &  978 & 252 & 63 \\
Observed               & 1,347,900 & 248,816 & 59,998 & 14,377 & 3587 &  898 & 217 & 48 \\
\hline\hline
& \multicolumn{8}{c|}{Muon channel} \\ \hline
& $\texttt{+}\geq\mathrm{0~jets} $ & $\texttt{+}\geq\mathrm{1~jets}$ & $\texttt{+}\geq\mathrm{2~jets}$ & $\texttt{+}\geq\mathrm{3~jets}$ & $\texttt{+}\geq\mathrm{4~jets}$ & $\texttt{+}\geq\mathrm{5~jets}$ & $\texttt{+}\geq\mathrm{6~jets}$  & $\texttt{+}\geq\mathrm{7~jets}$\\
\hline
\Ztomm\  [\%]              & 99.3 & 97.5 & 94.0 & 90.7 & 88.3 & 86.7 & 84.8 & 84.6\\
Top quark [\%]             & 0.2  & 1.1  & 3.6  & 6.0  & 7.7  & 8.1  & 8.7 & 7.7\\
Diboson   [\%]             & 0.2  & 0.7  & 1.6  & 2.4  & 3.4  & 4.5  & 5.9  & 7.0\\
$Z\to\tau^{+}\tau^{-}$ [\%] & <0.1 & <0.1 & <0.1 & <0.1 & <0.1 & <0.1 & <0.1 & <0.1\\
$W \to \mu\nu$   [\%]      & <0.1 & <0.1 & <0.1 & <0.1 & <0.1 & <0.1 & <0.1 & <0.1\\
Multijet         [\%]      & 0.3  & 0.6  & 0.9  & 0.9  & 0.7  & 0.7  & 0.7  & 0.7\\
\hline
Expected               & 1,693,000 & 300,600 & 71,230 & 17,740 & 4523 & 1187 &  307  & 76 \\ 
Observed               & 1,708,602 & 311,183 & 74,510 & 17,865 & 4387 & 1081 &  240  & 57\\
\hline\hline
\end{tabular}
\end{footnotesize}
\caption{Fraction of signal and background processes in \% in the final selection and expected and observed numbers of events for the various inclusive jet multiplicities considered in the electron (top) and muon (bottom) channels.}
\label{tab:zjets_sigbak}
\end{table}
  
\subsection{Correction factors and related systematic uncertainties}
\label{sec:uncertainties}
Some of the object and event selection efficiencies as well as the energy and momentum calibrations modelled by the simulation must be corrected with simulation-to-data correction factors to better match those observed in the data. These corrections and their corresponding uncertainties fall into the following two categories: dependent and not dependent on lepton flavour.

The corrections and uncertainties specific to each leptonic final state (\Ztoee and \Ztomm) are as follows:
\begin{itemize}
\item \textbf{Trigger:} The lepton trigger efficiency is estimated in simulation, with a separate data-driven analysis
performed to obtain the simulation-to-data trigger correction factors and their corresponding uncertainties~\cite{Aaboud:2016leb}.
\item \textbf{Lepton reconstruction, identification, and isolation:} The lepton selection efficiencies as determined from simulation are also corrected with simulation-to-data correction factors, with corresponding uncertainties~\cite{ATL-PHYS-PUB-2015-041,Aad:2016jkr}. 
\item \textbf{Energy, momentum scale/resolution:} Uncertainties in the lepton calibrations are estimated~\cite{Aad:2016jkr} because they can cause a change of acceptance because of migration of events across the \pT threshold and $m_{\ell\ell}$ boundaries.
\end{itemize}     

The corrections and uncertainties common to the electron and muon final states are as follows:
\begin{itemize}
\item \textbf{Jet energy scale and resolution}:
Uncertainties in the jet energy-scale calibration and resolution have a significant impact on the measurements, especially for the higher jet multiplicities.  The jet energy-scale calibration is based on 13~TeV simulation and on in situ corrections obtained from data~\cite{ATLAS-PHYS-PUB-2015-015}. The uncertainties are estimated using a decorrelation scheme, resulting in a  set of 19 independent parameters which cover all of the relevant calibration uncertainties. The jet energy scale is the dominant systematic uncertainty for all bins with at least one jet.
The jet energy-resolution uncertainty is derived by over-smearing the jet energy in the simulation and using the symmetrised variations as the uncertainty.
\item \textbf{Jet vertex tagger}: The modelling of the output variable from the jet vertex tagger is validated using data events where the \Zboson boson recoils against a jet. A percent-level correction is derived and its statistical and systematic uncertainties are used as additional uncertainties in the efficiency to select jets from the primary vertex~\cite{ATLAS-CONF-2014-018}.  
\item \textbf{Pile-up}: The imperfect modelling of the effects of pile-up leads to acceptance changes at the percent level for different jet multiplicities. To assess this uncertainty, the average number of interactions per bunch crossing $\langle \mu \rangle$ is varied in simulation so that the behaviour of variables sensitive to pile-up matches that observed in data. 
\item \textbf{Luminosity}: The cross sections have a 2.1\% uncertainty from the measurement of the integrated luminosity, which is derived, following a methodology similar to that detailed in Refs.~\cite{DAPR-2011-01,Aaboud:2016hhf}, from a calibration of the luminosity using $x$--$y$ beam-separation scans performed in August 2015. 
\end{itemize}

\section{Background estimation}
\label{sec:backgrounds}
Contributions from the electroweak (single boson and diboson) and top-quark (single top-quark and top-quark pair) components of the background are estimated using the Monte Carlo samples described in Section~\ref{sec:dataMC} with corresponding uncertainties as listed in Table~\ref{tab:MC}. Contributions from multijet events are evaluated with data-driven techniques as described below. A summary of the composition and relative importance of the backgrounds in the candidate \Zj events is given in Table~\ref{tab:zjets_sigbak}. The overall purity of the  \Zj\ selections (fraction of signal events in the final selection) ranges from 99\% in the inclusive sample to 80-85\% in the $\geq\mathrm{7~jets}$ bin.

\subsection{Top-quark and electroweak backgrounds}
\label{sec:ew_bkg}

The dominant contribution to the background at high jet multiplicities comes from $t\overline{t}$ production, with the subsequent leptonic decays of the \Wboson bosons originating from the top quarks and is evaluated from simulation.   An overall uncertainty of 6\%, corresponding to the PDF and scale variations on the theoretical predictions of the inclusive cross sections, is assigned (see Table~\ref{tab:MC}).  The $t\overline{t}$ background estimate is validated through a  cross-section measurement of $t\overline{t}$ production in the dilepton channel at  $\sqrt s = 13$~TeV~\cite{ATLAS-CONF-2015-065} as a function of the jet multiplicity, and the modelling of the additional parton radiation in $t\overline{t}$ events by \POWHEGPysix was found to be in good agreement with this measurement. In addition, a systematic uncertainty in the modelling of the shape of the distributions is derived by modifying the parton-shower intensity  in the nominal simulation sample and by comparing to the predictions from the alternative generator \MGaMCHpp (both listed in  Table~\ref{tab:MC}).
The small contribution from single-top-quark events is also estimated using \POWHEGPysix samples and assigned a 6\% uncertainty.

Diboson production in leptonic and semileptonic final states with at least two leptons of the same flavour constitutes a co-dominant background for high jet multiplicities (see Table~\ref{tab:zjets_sigbak}). The production of $WZ$ bosons in association with jets at $\sqrt s=13$~TeV was found to be well modelled by the \SHERPAtwoone generator~\cite{Aaboud:2016yus}. A 6\% uncertainty, again corresponding to PDF and scale variations on the predictions, is assessed.
Since in Ref.~\cite{Aaboud:2016yus} the measurement is limited by the statistical precision for dibosons $+\geq 4$~jets (resulting in $\geq 6$~hadronic jets for semileptonic diboson decays), an additional systematic uncertainty of 50\% in the normalisation of the diboson background is added for \Zsixjs.

Minor background contributions also arise from single-\Wboson-boson production decaying to leptonic final states and from single-\Zboson-boson production in the \Ztotautau process, both estimated with simulation and assigned a 5\% uncertainty (as given in Table~\ref{tab:MC}).

\subsection{Multijet background}
\label{sec:mj_bkg}

Background-enriched data control regions are used to estimate the multijet contribution in both the electron and muon channels. They are constructed by loosening the lepton identification and isolation requirements. Templates are built from the dilepton invariant mass distribution, a variable that shows discrimination between multijet background and other processes in regions of its kinematic range, but is largely uncorrelated with the variables used to build the multijet control regions. The templates are  subsequently normalised to events passing the \Zboson-boson signal selection. 

In the electron channel, the multijet templates are built for each jet multiplicity from  events with two same-charge leptons with no isolation requirement, whose identification criteria are looser than those of the signal selection, which the leptons must not satisfy. In the muon channel, the control region is similarly built from events with two leptons which are selected with looser identification requirements than the signal selection and also fail the nominal isolation requirement. In both cases, dedicated triggers better suited to this purpose are used to populate the templates. The small electroweak and top-quark contamination is  subtracted using simulated events. 

The normalisation of the multijet template is estimated with a log-likelihood fit to the measured dilepton invariant mass distribution for  the inclusive \Zboson selection, using  templates for  \Ztoll and for the electroweak and top-quark background derived from simulation. 
The fit is performed in the invariant mass windows of $52 < m_{ee} < 148$~GeV and $40 < m_{\mu\mu} < 80$~GeV for the electron and muon channels, respectively, in order to benefit from the larger multijet contribution in the mass sidebands. The normalisation of the multijet template is allowed to float freely while the remaining non-multijet templates are constrained to be within 6\% of the predicted cross sections for these processes as given in Table~\ref{tab:MC}. The multijet fractions are evaluated separately for each jet multiplicity, except for very high jet multiplicities where the templates are statistically limited, and so these fractions are taken from the estimates of the  $\geq\mathrm{5~jets}$ and $\geq\mathrm{4~jets}$ bins in the electron and muon channels, respectively.

The systematic uncertainties on the  multijet background are derived by varying the mass range and bin width of the nominal fit, using the lepton transverse impact parameter $d_{0}$ as the fitting variable instead of the invariant mass, using alternative simulation samples for the templates, allowing the normalisations of the non-multijet components to vary independently or within a wider range, and varying the lepton resolution and energy/momentum scales. In addition, given the multiple sources of multijet background in the electron channel, an alternative template is constructed by requiring that the electrons fail to meet an isolation criterion instead of failing to meet the nominal signal selection electron identification criterion. 

The resulting estimated multijet fractions in each jet multiplicity bin are given in Table~\ref{tab:zjets_sigbak}. Their corresponding total uncertainties are dominated by their systematic components. These systematic components are approximately 70\%  of the multijet fraction as estimated in the electron and muon channels.

\section{Kinematic distributions}
\label{sec:kinematics}
\begin{figure}[p]
\centering
\includegraphics[width=0.49\textwidth]{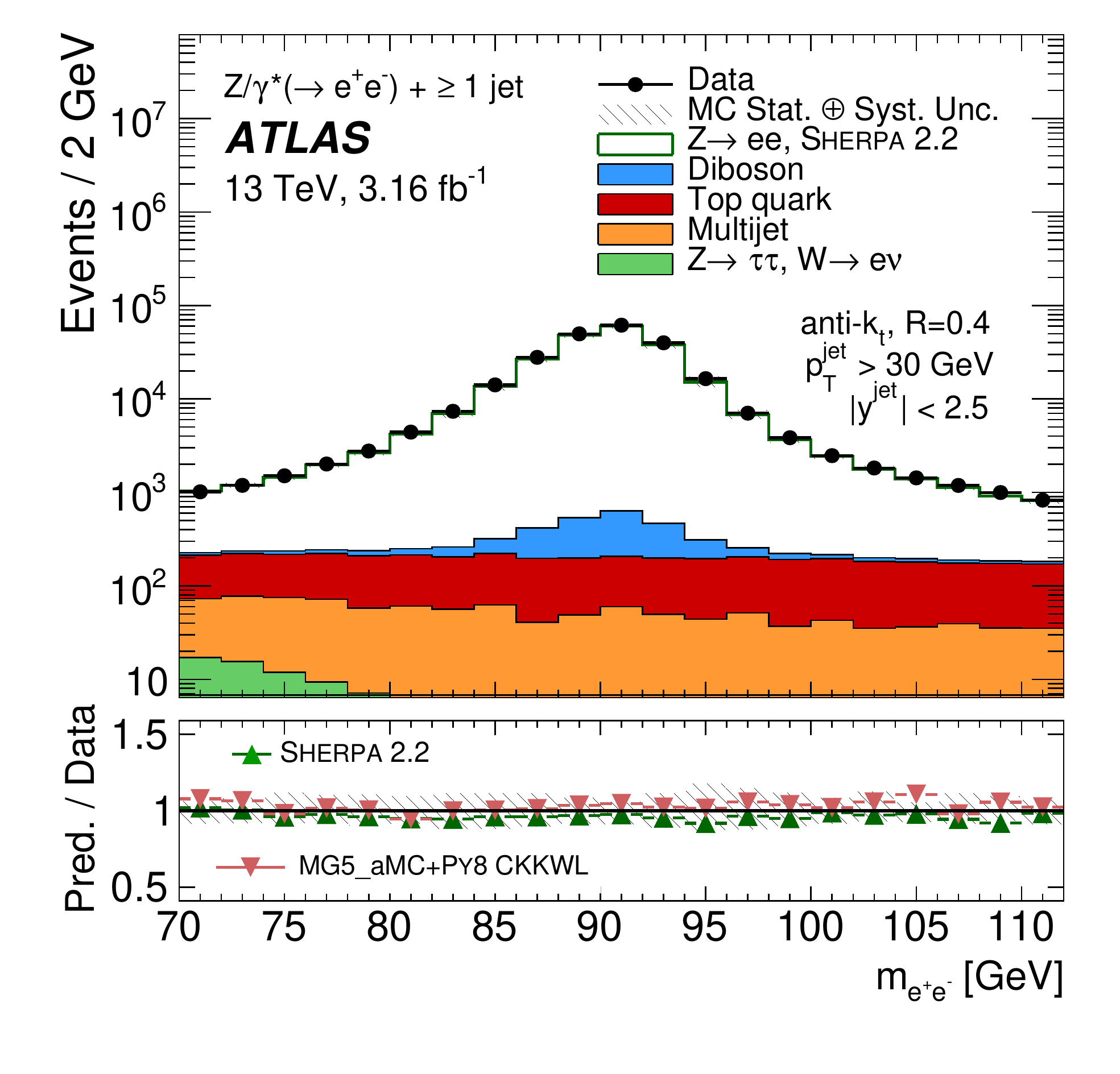}
\includegraphics[width=0.49\textwidth]{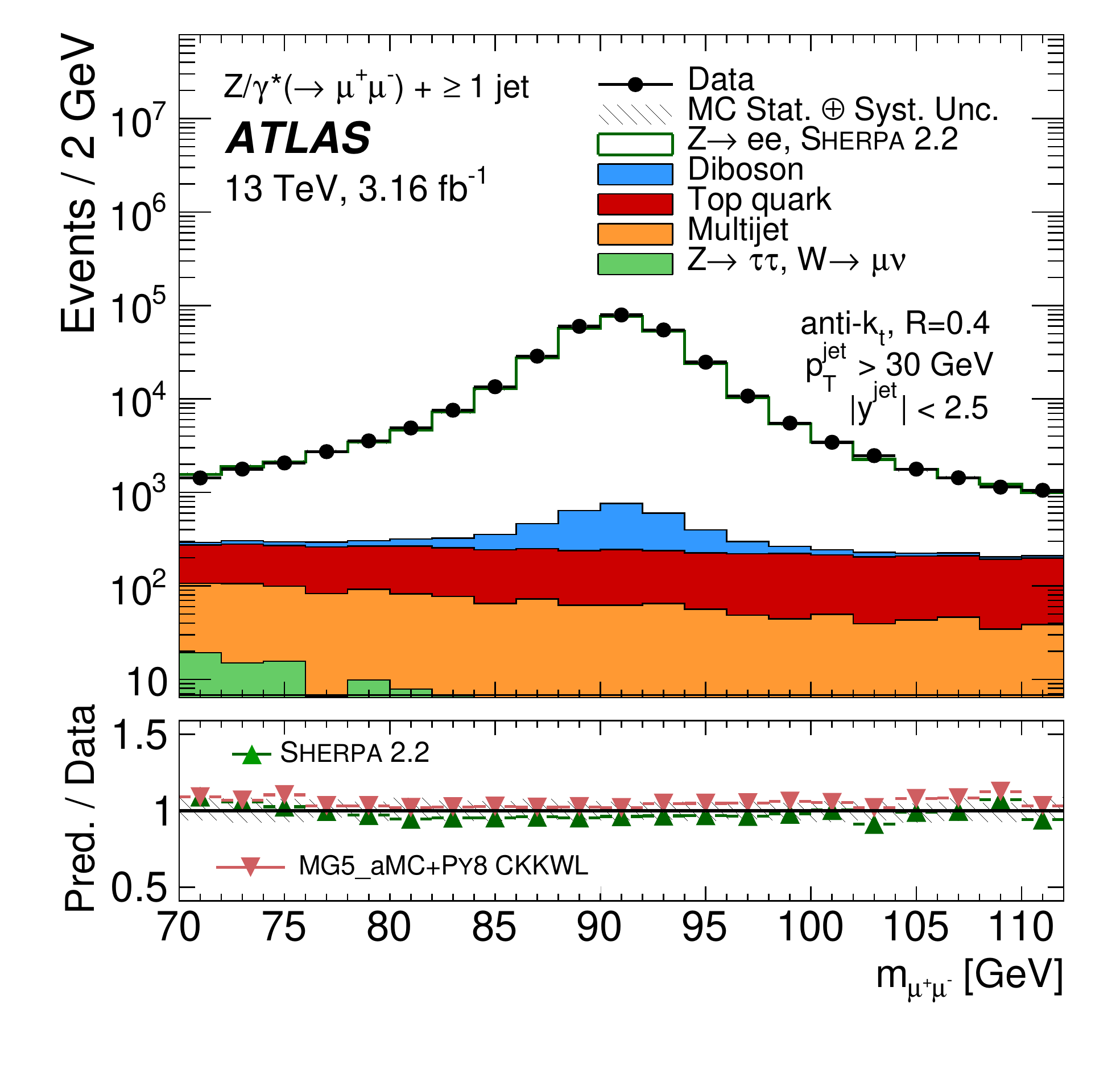}
\includegraphics[width=0.49\textwidth]{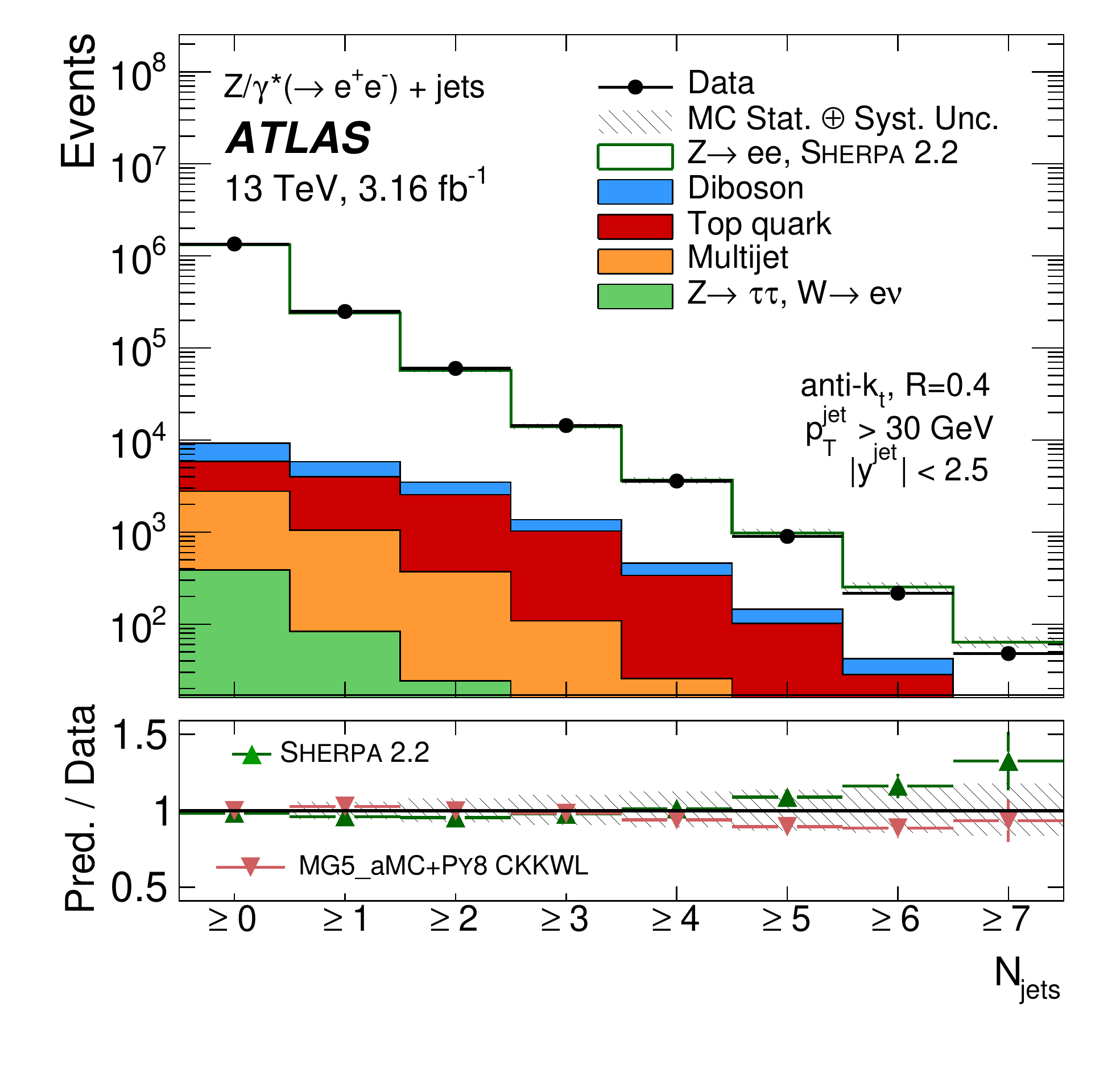}
\includegraphics[width=0.49\textwidth]{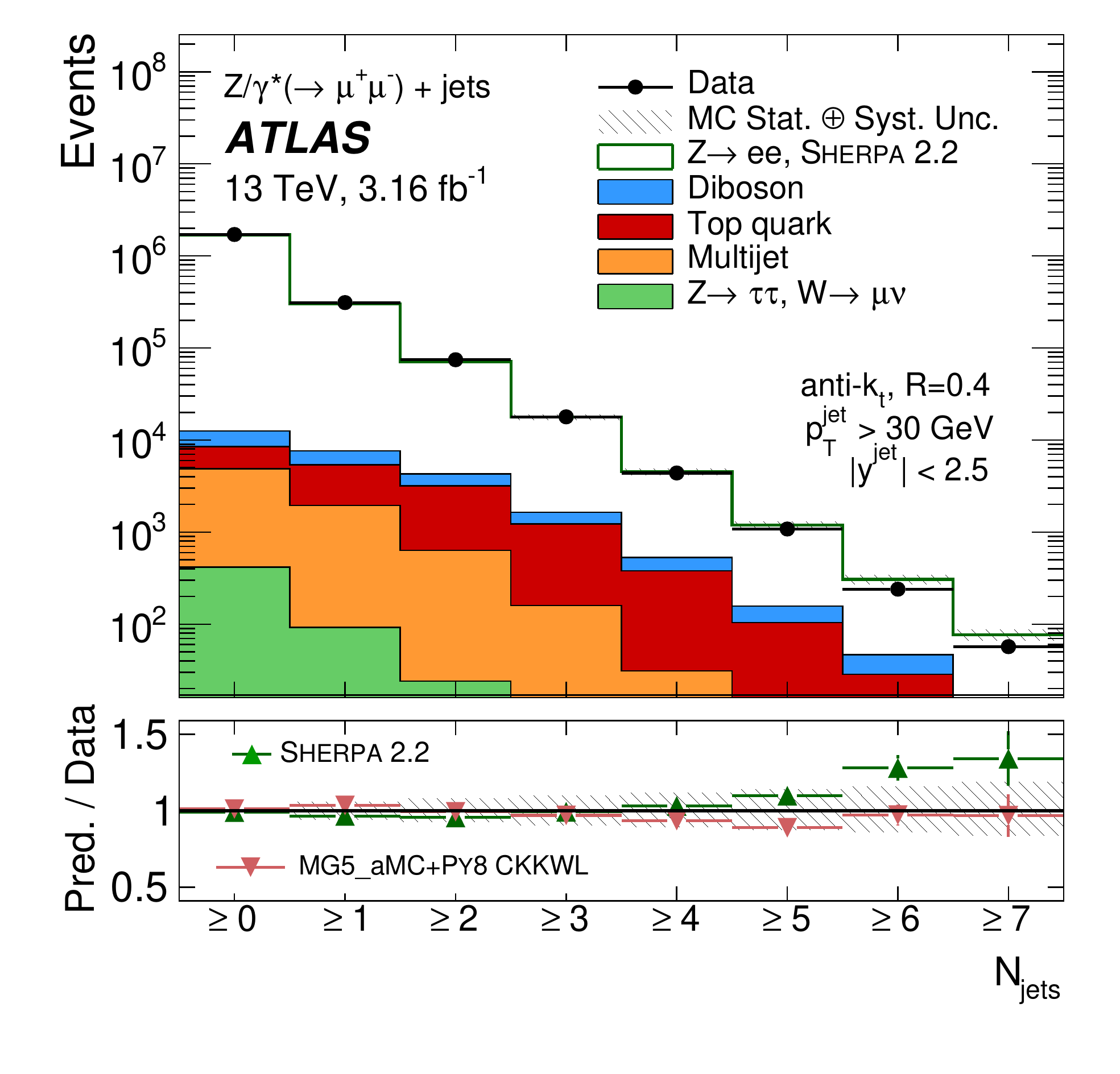}
\caption{Dilepton invariant mass for \Zonej (top) and inclusive jet multiplicity (bottom) in the \Ztoeej (left) and the \Ztommj (right) channels.  All backgrounds and the signal samples are stacked to produce the figures.
Systematic uncertainties for the signal and background distributions are combined in the hatched band, and the statistical uncertainty is shown on the data points. The uncertainty in the luminosity and the theory uncertainty in the signal prediction are not included in the uncertainty band.}
\label{fig:ZJetMass123j}
\end{figure}

The level of agreement between data and predictions is evaluated from the comparison of kinematic distributions.
Figure~\ref{fig:ZJetMass123j}, which presents the dilepton mass for the \Zonej topology and the inclusive jet multiplicity, shows how well the  
\SHERPAtwotwo and \MGaMCPyCKKWL predictions agree with data. 
The uncertainty bands shown in these distributions include the statistical uncertainties due to the simulation sample sizes, the event-selection uncertainties described in Section~\ref{sec:uncertainties} (omitting the common 2.1\% luminosity uncertainty), and the  background normalisation uncertainties described in Section~\ref{sec:backgrounds}.

\section{Unfolding of detector effects}
\label{sec:unfolding}
The cross-section measurements presented in this paper are performed within the fiducial acceptance region defined by the following requirements:
\begin{itemize}
\item $\pT^{\ell}>25$~GeV, $|\eta^{\ell}|<2.5$
\item $\pTj > 30$~GeV, $|\yj|<2.5$  
\item $\Delta R(\ell,\text{jet}) > 0.4$
\item $71<m_{\ell\ell}<111$~GeV.
\end{itemize}
The  cross sections are defined at particle (``truth'') level, corresponding to dressed electrons and muons from the \Zboson bosons. The particle level also includes jets clustered  using the anti-$k_t$ algorithm~\cite{Cacciari:2008} with radius parameter $R=0.4$ for final-state particles with decay length $c\tau>10$~mm, excluding the dressed \Zboson-boson decay products.

The fiducial cross sections are estimated from the reconstructed kinematic observables: jet multiplicity, \pTj for different jet multiplicities, \yj, \Dphijj,  \mjj, and \HT, for events that pass the selection described in Section~\ref{sec:selection}. The expected background components as described in Section~\ref{sec:backgrounds} are subtracted from the distributions in data.  A variable-width binning of these observables is used, such that the purity is at least 50\% in each bin and the size of the statistical uncertainty in most of the bins remains below 10\%. 

\begin{figure}[t]
\begin{center}
\vspace{10mm}
\includegraphics[width=0.49\textwidth]{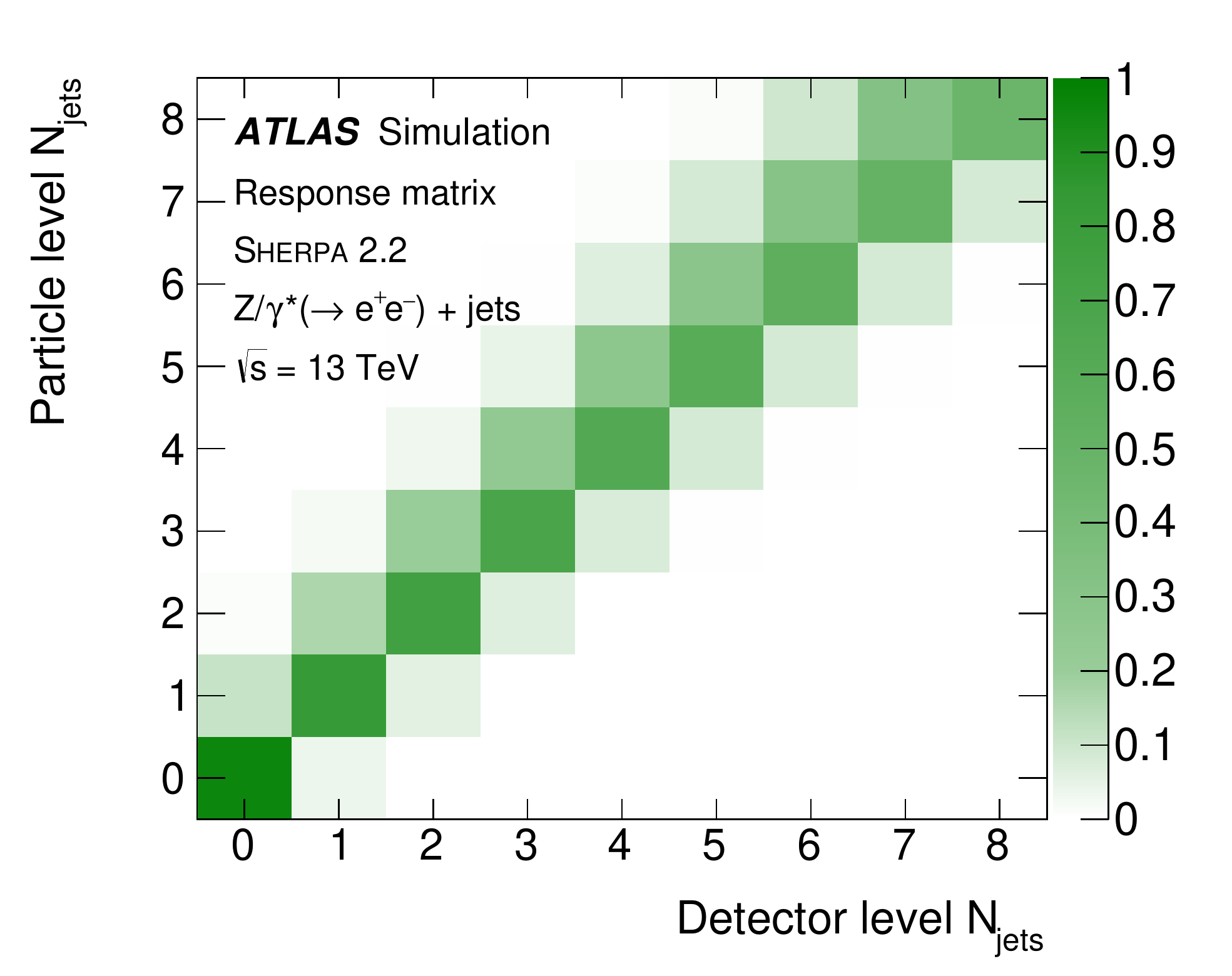}
\includegraphics[width=0.49\textwidth]{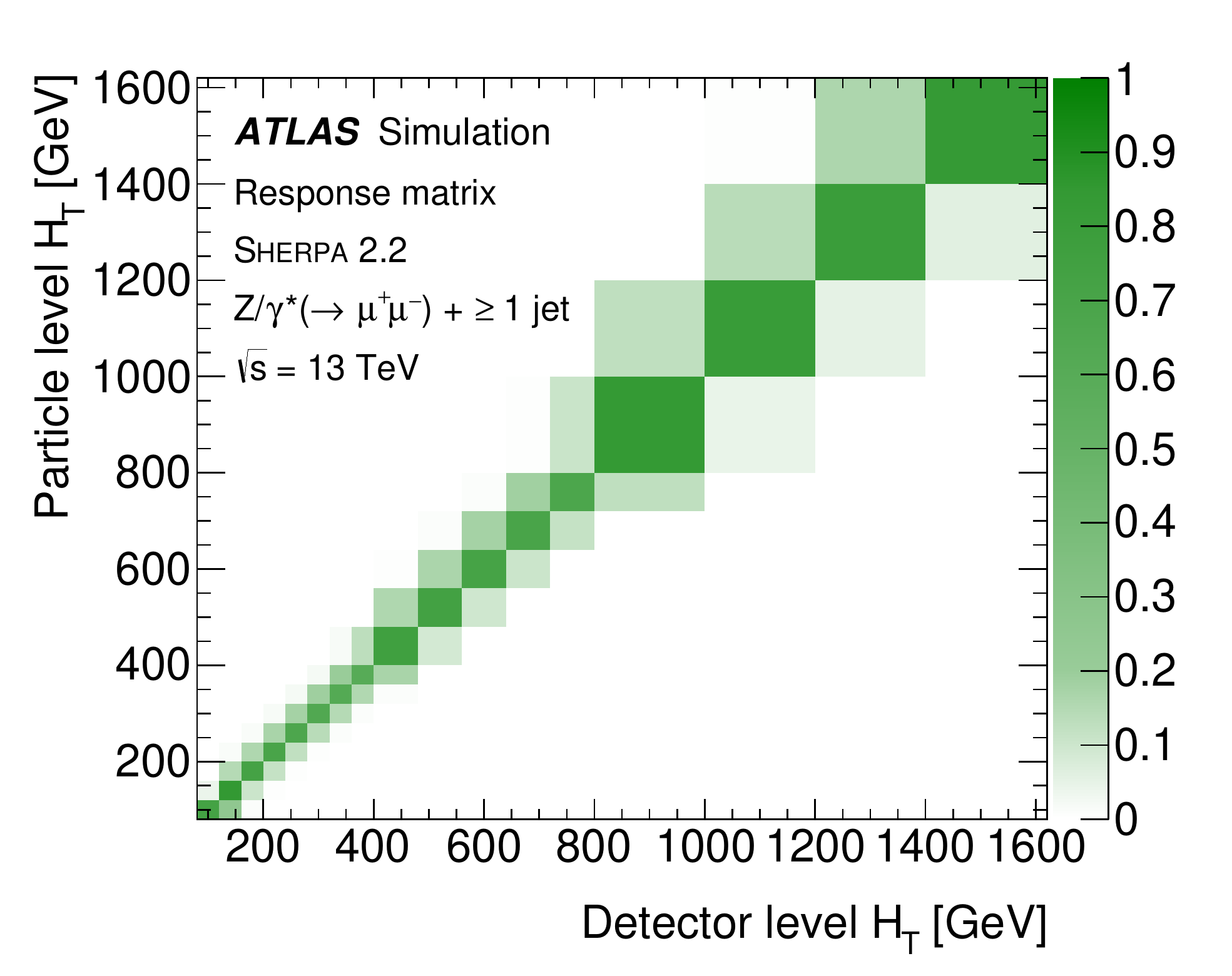}\\
\caption{Response matrices corresponding to the exclusive jet multiplicity for \Zj events in the electron channel (left)  and to the \HT for \Zonej events in the  muon channel (right). The sum of the entries in each row is normalised to unity. Both matrices are obtained from \SHERPAtwotwo.}
\label{fig:matrix_sh_1}
\end{center}
\end{figure}

An iterative Bayesian unfolding technique~\cite{D'Agostini:1994zf}, as implemented in the RooUnfold package~\cite{RooUnfold}, is used to unfold the measurements to the particle level, thereby accounting for detector effects related to inefficiencies, resolution, and systematic biases in the central values of the kinematic variables describing both the leptons and the jets. The iterative unfolding technique updates the initial estimators for the generated (``truth'') distribution in consecutive steps, using Bayes' theorem in each iteration to derive an unfolding matrix from the initial response matrix (which relates truth and reconstructed distributions of given observables) and the current truth estimator.

The response  matrices are constructed using the \SHERPAtwotwo \Ztollj samples. \SHERPAtwotwo is also used to derive the initial truth estimator. In order to enter the response matrix, events must pass the \Zboson-boson selection at generator level and at detector level and contain the  number of jets required by the preselection for a given observable at both generator and detector level. Reconstructed jets are required to match the corresponding generator-level jets within a cone of size $\Delta R=0.4$ for all distributions except global quantities such as the jet multiplicity and \HT. A given bin $(i,j)$ in the response matrix therefore corresponds to the probability that a true jet object in bin $j$ is reconstructed in bin $i$ of the distribution. 
Figure~\ref{fig:matrix_sh_1} illustrates two examples of response matrices.
The resulting ratios of detector-level  to truth-level event yields are typically 0.65 and 0.8 for the electron and muon channels, respectively.

The background-subtracted data are corrected for the expected fraction of events with reconstructed objects unmatched to any generator object before entering the iterative unfolding.
The number of iterations used for the iterative unfolding of each distribution (two) is chosen by unfolding the \SHERPAtwotwo samples reweighted to data and comparing to the generated reweighted distribution.
The unfolded event yields are divided by the integrated luminosity of the data sample and the bin width of the distribution in question to provide the final fiducial cross sections. The final result is given by

\begin{equation}
\sigma_i  = \frac{1}{\epsilon_i  L} \sum_j U_{ij} N_j^{\text{data}} \left(1-f_j^{\text{unmatched}}\right),
\end{equation}

where $L$ is the integrated luminosity, $\epsilon_i$ is the reconstruction efficiency for truth bin $i$, $N_j^{\text{data}}$ corresponds to the number of events observed in data in reconstructed bin $j$ and $f_j^{\text{unmatched}}$ is its fraction of unmatched events calculated from simulation, and $U_{ij}$ is the unfolding matrix calculated after two iterations, using the updated prior from the first iteration and the response matrix.

\subsection{Systematic uncertainties associated with the unfolding procedure}

The limited size of a simulation sample can create biases in the distributions. Systematic uncertainties  account for possible residual biases in the unfolding procedure due to, e.g. modelling of the hadronisation in the simulation, migrations into other kinematic distributions not explicitly part of the unfolding, or the finite bin width used in each distribution.  The following uncertainties arise from the unfolding procedure.

\begin{itemize}

\item The statistical uncertainties of the response matrices derived from \SHERPAtwotwo are propagated to the unfolded cross sections with a toy simulation method. A total of 5000 ensembles (pseudo-experiments) of unfolded samples are generated. For each sample, the number of reconstructed events in each bin is generated randomly according to a Gaussian distribution, where the mean is the nominal number of events before unfolding and the width is its corresponding statistical uncertainty. 
Unfolding is performed for each ensemble. The widths of resulting distributions are taken as a systematic uncertainty of the unfolding.

\item  The \SHERPAtwotwo samples are reweighted at  generator level, such that the distribution of the leading jet \pt at detector level matches that observed in the data. The modified \SHERPAtwotwo samples are then used to unfold the data again and the variations in the resulting cross sections are used to derive a systematic uncertainty.  

\item An additional check is performed by unfolding  reconstructed \MGaMCPyCKKWL events using \SHERPAtwotwo response matrices. The residual non-closure is accounted for by an additional flat uncertainty of  3\% for all distributions.

\end{itemize}

\section{Results}
\label{sec:results}

The measured cross sections, presented in Section~\ref{sec:results_channels}, are calculated in the electron and muon channels separately and the compatibility of the results of the two channels is evaluated.  In order to improve the precision of the measurement, these results are then combined, taking into account the correlations of the systematic uncertainties. 
The comparisons of the combined results to the predictions are presented in Section~\ref{sec:comparison}.

\subsection{Results in the individual channels and the combination}
\label{sec:results_channels}

The fiducial cross-section measurements in the  \Ztoeej and  \Ztommj channels as a function of the inclusive jet multiplicities are presented in Table~\ref{tab:zjets_njet}. The data statistical uncertainties are propagated through the unfolding by using pseudo-experiments. As mentioned in Section~\ref{sec:unfolding}, the systematic uncertainties  are propagated through the unfolding via the migration matrices and via the variation of the subtracted background.  Table~\ref{tab:systot_ell} shows the resulting total relative statistical and systematic uncertainties as well as the systematic components (lepton trigger, lepton selection, jet energy scale and resolution, jet vertex tagging, pile-up, luminosity (all described in Section~\ref{sec:uncertainties})), unfolding (described in Section~\ref{sec:unfolding}), and background (described in Section~\ref{sec:backgrounds}) as a function of the inclusive jet multiplicity, presented separately for the electron and muon channels. 
The jet energy scale is the dominant systematic uncertainty for all bins with at least one jet.

\begin{table}[t]
\centering
\begin{tabular}{|c|r@{\,}@{$\pm$}@{\;}r@{\,}@{$\pm$}@{\;}r@{\,}@{$\pm$}@{\;}r|r@{\,}@{$\pm$}@{\;}r@{\,}@{$\pm$}@{\;}r@{\,}@{$\pm$}@{\;}r|}
\hline\hline
Jet & \multicolumn{8}{c|}{Measured cross section $\pm$\,(stat.)\,$\pm$\,(syst.)\,$\pm$\,(lumi.)   [pb]} \\
\cline{2-9}\cline{2-9}
multiplicity & \multicolumn{4}{c|}{$Z \to ee$} & \multicolumn{4}{c|}{$Z \to \mu\mu$}\\ 
\hline\hline
$\geq\mathrm{0~jets}$ & \numRP{742.6131254}{0} & \numRP{0.6560087}{0} & \numRP{24.1232395}{0} & \numRP{15.6790915}{0}
                      & \numRP{738.4631887}{0} & \numRP{0.5785262}{0} & \numRP{23.4250717}{0} & \numRP{15.5881443}{0} \\
$\geq\mathrm{1~jets}$ & \numRP{116.6183205}{1} & \numRP{0.2697211}{1} &  \numRP{9.8539801}{1} &  \numRP{2.5056422}{1}
                      & \numRP{115.6865698}{1} & \numRP{0.2400960}{1} &  \numRP{9.7178793}{1} &  \numRP{2.4838357}{1} \\
$\geq\mathrm{2~jets}$ &  \numRP{27.0842355}{1} & \numRP{0.1258711}{1} &  \numRP{2.9171276}{1} &  \numRP{0.6042993}{1}
                      &  \numRP{26.9592355}{1} & \numRP{0.1120992}{1} &  \numRP{2.8405800}{1} &  \numRP{0.5999904}{1} \\
$\geq\mathrm{3~jets}$ &   \numRP{6.2010092}{2} & \numRP{0.0581154}{2} &  \numRP{0.8179384}{2} &  \numRP{0.1442643}{2}
                      &   \numRP{6.2192955}{2} & \numRP{0.0525592}{2} &  \numRP{0.8330854}{2} &  \numRP{0.1440473}{2} \\
$\geq\mathrm{4~jets}$ &   \numRP{1.4870271}{2} & \numRP{0.0277629}{2} &  \numRP{0.2341416}{2} &  \numRP{0.0360207}{2}
                      &   \numRP{1.4819739}{2} & \numRP{0.0251189}{2} &  \numRP{0.2338763}{2} &  \numRP{0.0355780}{2} \\
$\geq\mathrm{5~jets}$ &   \numRP{0.3566019}{3} & \numRP{0.0132320}{3} &  \numRP{0.0690235}{3} &  \numRP{0.0089792}{3}
                      &   \numRP{0.3544801}{3} & \numRP{0.0119510}{3} &  \numRP{0.0682254}{3} &  \numRP{0.0087775}{3} \\
$\geq\mathrm{6~jets}$ &   \numRP{0.0817674}{3} & \numRP{0.0062634}{3} &  \numRP{0.0191435}{3} &  \numRP{0.0021483}{3}
                      &   \numRP{0.0755170}{3} & \numRP{0.0054169}{3} &  \numRP{0.0185040}{3} &  \numRP{0.0019692}{3} \\
$\geq\mathrm{7~jets}$ &   \numRP{0.0179856}{4} & \numRP{0.0028686}{4} &  \numRP{0.0051447}{4} &  \numRP{0.0004971}{4}
                      &   \numRP{0.0165519}{4} & \numRP{0.0026949}{4} &  \numRP{0.0059834}{4} &  \numRP{0.0004460}{4} \\
\hline\hline
\end{tabular}
\caption{Measured fiducial cross sections in the electron and muon channels for successive inclusive jet multiplicities. 
The total statistical and systematic uncertainties are given, along with the uncertainty in the luminosity.}
\label{tab:zjets_njet}
\end{table}

\begin{table}[t]
\centering
\begin{footnotesize}
\begin{tabular}{|l|r|r|r|r|r|r|r|r|}
\hline\hline
\multicolumn{9}{|c|} {Relative uncertainty in $\sigma(\ZNj)$ [\%] }\\\hline
& \multicolumn{8}{c|}{\Ztoee}  \\\hline  
Systematic source &
$\texttt{+}\geq\text{0~jets}$ &
$\texttt{+}\geq\text{1~jets}$ &
$\texttt{+}\geq\text{2~jets}$ &
$\texttt{+}\geq\text{3~jets}$ &
$\texttt{+}\geq\text{4~jets}$ &
$\texttt{+}\geq\text{5~jets}$ &
$\texttt{+}\geq\text{6~jets}$ &
$\texttt{+}\geq\text{7~jets}$ \\\hline  
Electron trigger       & \numRP{0.0719}{1} & \numRP{0.0757}{1} & \numRP{0.0727}{1} & \numRP{0.1571}{1} & \numRP{0.2140}{1} & \numRP{0.1635}{1} & \numRP{0.2540}{1} & \numRP{0.2757}{1} \\
Electron selection     & \numRP{1.1546}{1} & \numRP{1.5635}{1} & \numRP{1.8100}{1} & \numRP{1.8707}{1} & \numRP{2.2542}{1} & \numRP{2.6626}{1} & \numRP{2.9421}{1} & \numRP{3.8011}{1} \\
Jet energy scale       &            $<0.1$ & \numRP{6.6330}{1} & \numRP{9.1967}{1} & \numRP{11.5040}{1} & \numRP{13.7779}{1} & \numRP{17.2934}{1} & \numRP{20.5630}{1} & \numRP{23.6686}{1} \\
Jet energy resolution  &            $<0.1$ & \numRP{3.7428}{1} & \numRP{3.7497}{1} & \numRP{4.4201}{1} & \numRP{5.2532}{1} & \numRP{5.1719}{1} & \numRP{6.1797}{1} & \numRP{7.2990}{1} \\
Jet vertex tagger      &            $<0.1$ & \numRP{1.2748}{1} & \numRP{2.1030}{1} & \numRP{2.8384}{1} & \numRP{3.6007}{1} & \numRP{4.5365}{1} & \numRP{5.5114}{1} & \numRP{6.3037}{1} \\
Pile-up                & \numRP{0.4262}{1} & \numRP{0.1924}{1} & \numRP{0.0834}{1} & \numRP{0.1937}{1} & \numRP{0.1990}{1} & \numRP{0.0680}{1} & \numRP{0.4433}{1} & \numRP{0.7767}{1} \\
Luminosity             & \numRP{2.1113}{1} & \numRP{2.1486}{1} & \numRP{2.2312}{1} & \numRP{2.3265}{1} & \numRP{2.4223}{1} & \numRP{2.5180}{1} & \numRP{2.6274}{1} & \numRP{2.7638}{1} \\
Unfolding              & \numRP{3.0020}{1} & \numRP{3.0150}{1} & \numRP{3.0300}{1} & \numRP{3.0278}{1} & \numRP{3.0444}{1} & \numRP{3.0848}{1} & \numRP{3.1107}{1} & \numRP{3.2463}{1} \\
Background             & \numRP{0.1288}{1} & \numRP{0.3269}{1} & \numRP{0.5677}{1} & \numRP{1.0406}{1} & \numRP{1.6342}{1} & \numRP{3.2801}{1} & \numRP{5.9866}{1} & \numRP{11.6266}{1} \\
\hline
Total syst. uncertainty      & \numRP{3.8744}{1} & \numRP{8.7194}{1} & \numRP{10.9994}{1} & \numRP{13.3942}{1} & \numRP{15.9309}{1} & \numRP{19.5191}{1} & \numRP{23.5594}{1} & \numRP{28.7378}{1} \\
\hline
Stat. uncertainty      & \numRP{0.0883}{1} & \numRP{0.2313}{1} & \numRP{0.4647}{1} & \numRP{0.9372}{1} & \numRP{1.8670}{1} & \numRP{3.7106}{1} & \numRP{7.6600}{1} & \numRP{15.9497}{1} \\
\hline\hline
& \multicolumn{8}{c|}{\Ztomm}  \\\hline  
Systematic source &
$\texttt{+}\geq\text{0~jets}$ &
$\texttt{+}\geq\text{1~jets}$ &
$\texttt{+}\geq\text{2~jets}$ &
$\texttt{+}\geq\text{3~jets}$ &
$\texttt{+}\geq\text{4~jets}$ &
$\texttt{+}\geq\text{5~jets}$ &
$\texttt{+}\geq\text{6~jets}$ &
$\texttt{+}\geq\text{7~jets}$ \\\hline  
Muon trigger           & \numRP{0.4313}{1} & \numRP{0.4553}{1} & \numRP{0.3933}{1} & \numRP{0.4764}{1} & \numRP{0.3796}{1} & \numRP{0.4663}{1} & \numRP{0.8860}{1} & \numRP{0.6118}{1} \\
Muon selection         & \numRP{0.8130}{1} & \numRP{0.8879}{1} & \numRP{0.9501}{1} & \numRP{1.0017}{1} & \numRP{1.0252}{1} & \numRP{1.5282}{1} & \numRP{4.2222}{1} & \numRP{16.5887}{1} \\
Jet energy scale       &            $<0.1$ & \numRP{6.7575}{1} & \numRP{9.1080}{1} & \numRP{11.8950}{1} & \numRP{14.0308}{1} & \numRP{16.9759}{1} & \numRP{20.9403}{1} & \numRP{23.6833}{1} \\
Jet energy resolution  &            $<0.1$ & \numRP{3.6083}{1} & \numRP{3.6004}{1} & \numRP{4.1285}{1} & \numRP{5.0434}{1} & \numRP{5.8645}{1} & \numRP{6.1655}{1} & \numRP{9.3093}{1} \\
Jet vertex tagger      &            $<0.1$ & \numRP{1.2623}{1} & \numRP{2.1000}{1} & \numRP{3.0523}{1} & \numRP{3.5924}{1} & \numRP{4.3715}{1} & \numRP{5.6099}{1} & \numRP{6.5514}{1} \\
Pile-up                & \numRP{0.4245}{1} & \numRP{0.0649}{1} & $<0.1$ & \numRP{0.2644}{1} & \numRP{0.5164}{1} & \numRP{0.0785}{1} & \numRP{0.4459}{1} & \numRP{0.9438}{1} \\
Luminosity             & \numRP{2.1109}{1} & \numRP{2.1470}{1} & \numRP{2.2255}{1} & \numRP{2.3161}{1} & \numRP{2.4007}{1} & \numRP{2.4762}{1} & \numRP{2.6076}{1} & \numRP{2.6945}{1} \\
Unfolding              & \numRP{3.0014}{1} & \numRP{3.0108}{1} & \numRP{3.0197}{1} & \numRP{3.0296}{1} & \numRP{3.0306}{1} & \numRP{3.0625}{1} & \numRP{3.0900}{1} & \numRP{3.2130}{1} \\
Background             & \numRP{0.1572}{1} & \numRP{0.4147}{1} & \numRP{0.6419}{1} & \numRP{0.9021}{1} & \numRP{1.6971}{1} & \numRP{4.0293}{1} & \numRP{7.4036}{1} & \numRP{12.8994}{1} \\
\hline
Total syst. uncertainty      & \numRP{3.8104}{1} & \numRP{8.6704}{1} & \numRP{10.7691}{1} & \numRP{13.5939}{1} & \numRP{15.9630}{1} & \numRP{19.4053}{1} & \numRP{24.6417}{1} & \numRP{36.2525}{1} \\
\hline
Stat. uncertainty      & \numRP{0.0783}{1} & \numRP{0.2075}{1} & \numRP{0.4158}{1} & \numRP{0.8451}{1} & \numRP{1.6950}{1} & \numRP{3.3714}{1} & \numRP{7.1731}{1} & \numRP{16.2815}{1} \\
\hline\hline
\end{tabular}
\end{footnotesize}
\caption{Relative statistical and systematic uncertainties (in \%) in the measured cross sections of $Z+\text{jets}$ production for successive inclusive jet multiplicities in the electron (top) and muon (bottom) channels.} \label{tab:systot_ell}
\end{table}

\begin{figure}[p]
\begin{center}
\includegraphics[width=0.48\textwidth]{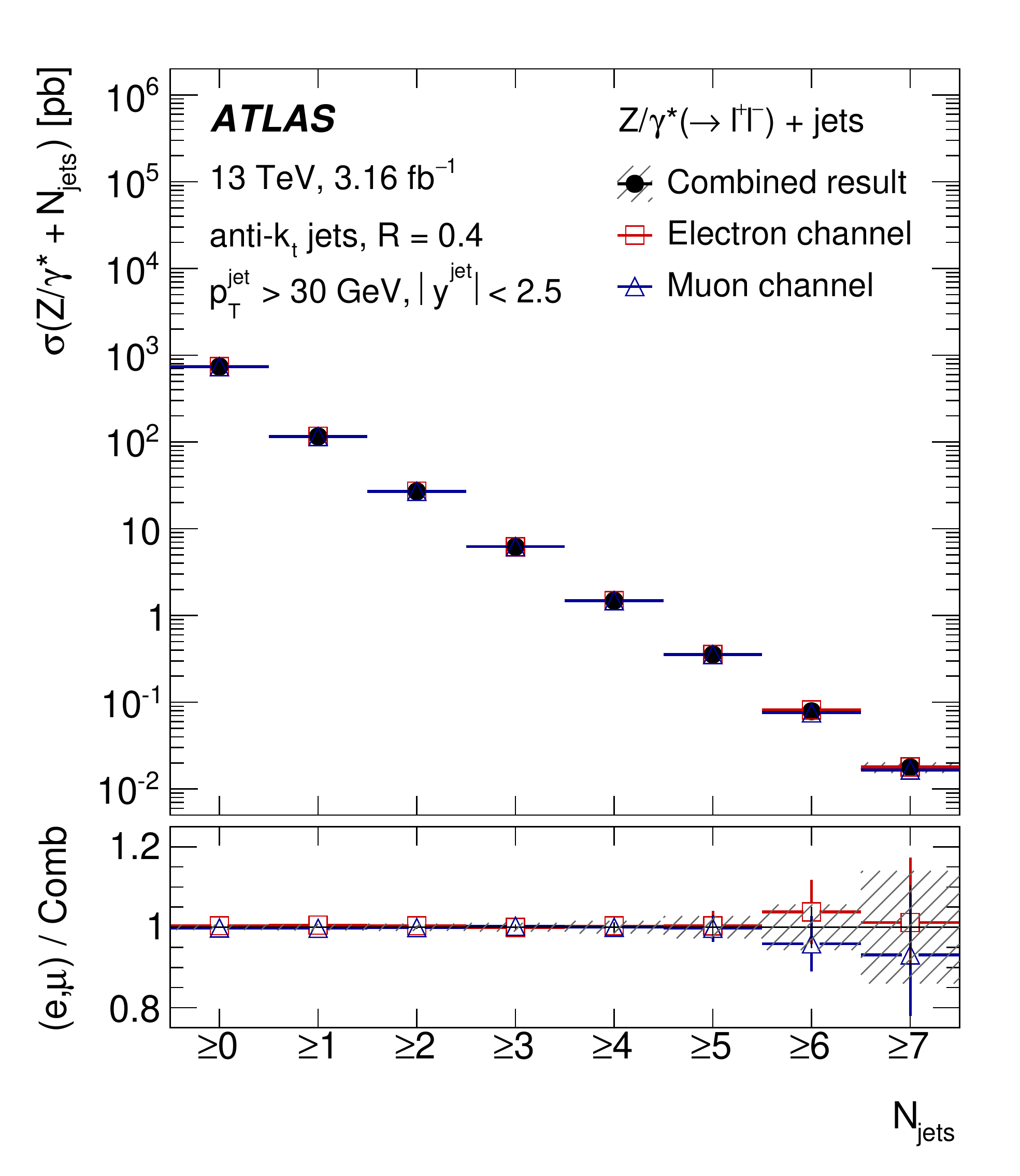}
\includegraphics[width=0.48\textwidth]{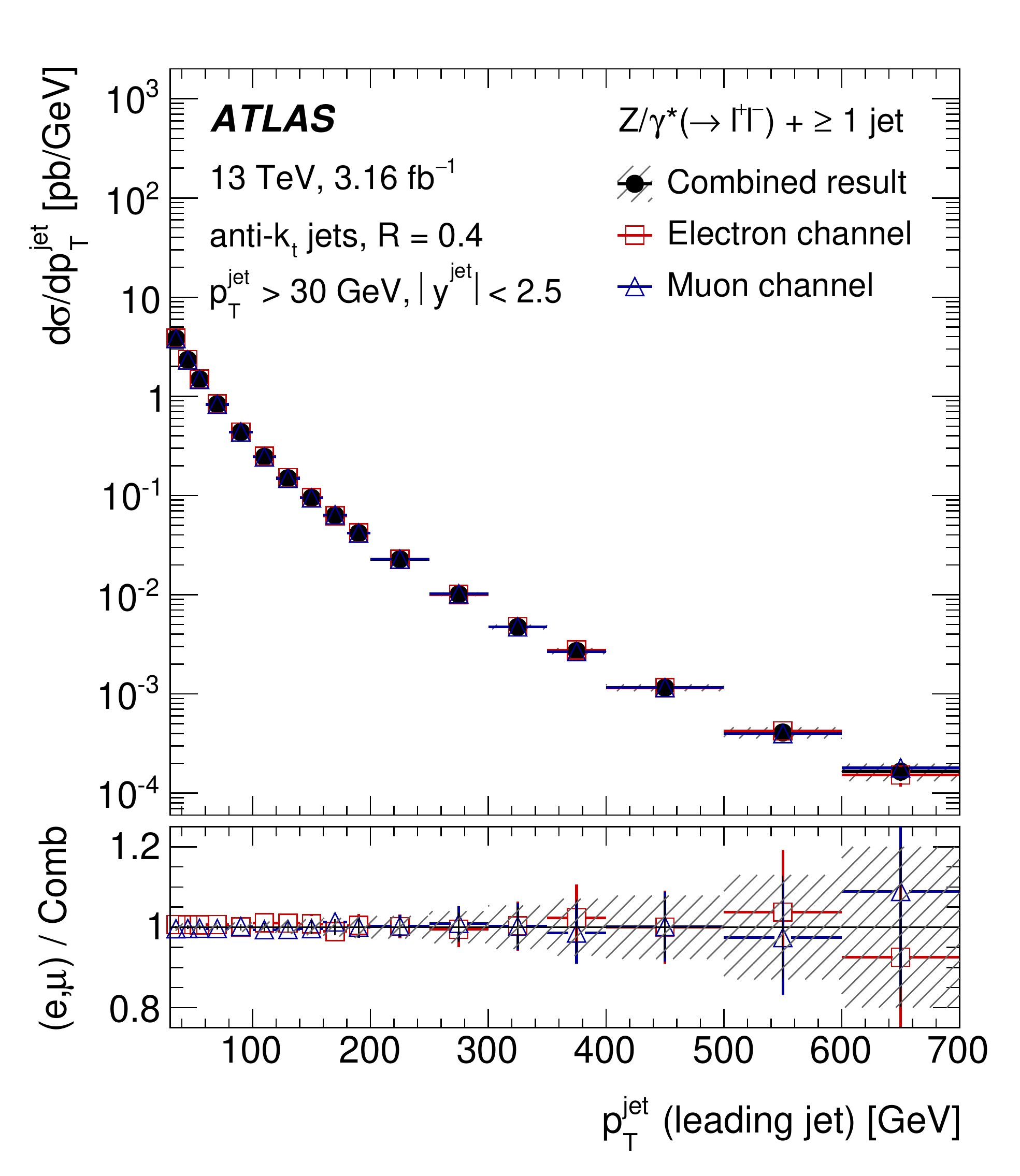}
\caption{Measured fiducial cross section as a function of the inclusive jet multiplicity (left) and the leading jet \pt\ for inclusive \Zonej events (right) in the electron and the muon channels and compared to their combined value. The ratios of the two measurements to the combined results are also shown in the bottom panels. 
The error bars indicate the statistical uncertainty, and the hatched bands 
the statistical and the flavour-uncorrelated systematic uncertainties of the combined result, added in quadrature.}
\label{fig:comp_Nin}
\end{center}
\end{figure}

Figure~\ref{fig:comp_Nin} shows a comparison of the electron and muon channels for the measured fiducial cross section as a function of the inclusive jet multiplicity and of the leading jet \pt\ for inclusive \Zonej events. This figure demonstrates that the results in the electron and muon channels are compatible and hence can be combined to improve the precision of the measurement. This figure also shows the result of this combination described below.

\clearpage

The results from the electron and muon channels are combined at dressed level for each distribution separately: inclusive and exclusive jet multiplicities, ratio for successive inclusive jet multiplicities, leading jet \pT for \Zonefourj events and jet \pT for exclusive  \Zexclj events, leading jet rapidity for inclusive \Zonej events, \HT,  \Dphijj, and \mjj. A $\chi^2$ function whose sum runs over all measurement sets (electrons and muons), measurement points, and some of the uncertainty sources, is used for the combination~\cite{Glazov:2005rn,Aaron:2009bp} and distinguishes between bin-to-bin correlated and uncorrelated sources of uncertainties, the latter comprising the statistical uncertainty of the data and the statistical unfolding uncertainty.  Uncertainties specific to the lepton flavour and to the background are included in the $\chi^2$ function, while the remaining, flavour-uncorrelated,  systematic uncertainties related to jets, pile-up, luminosity, and unfolding are averaged after the combination.

\begin{figure}[t]
\begin{center}
\includegraphics[width=0.48\textwidth]{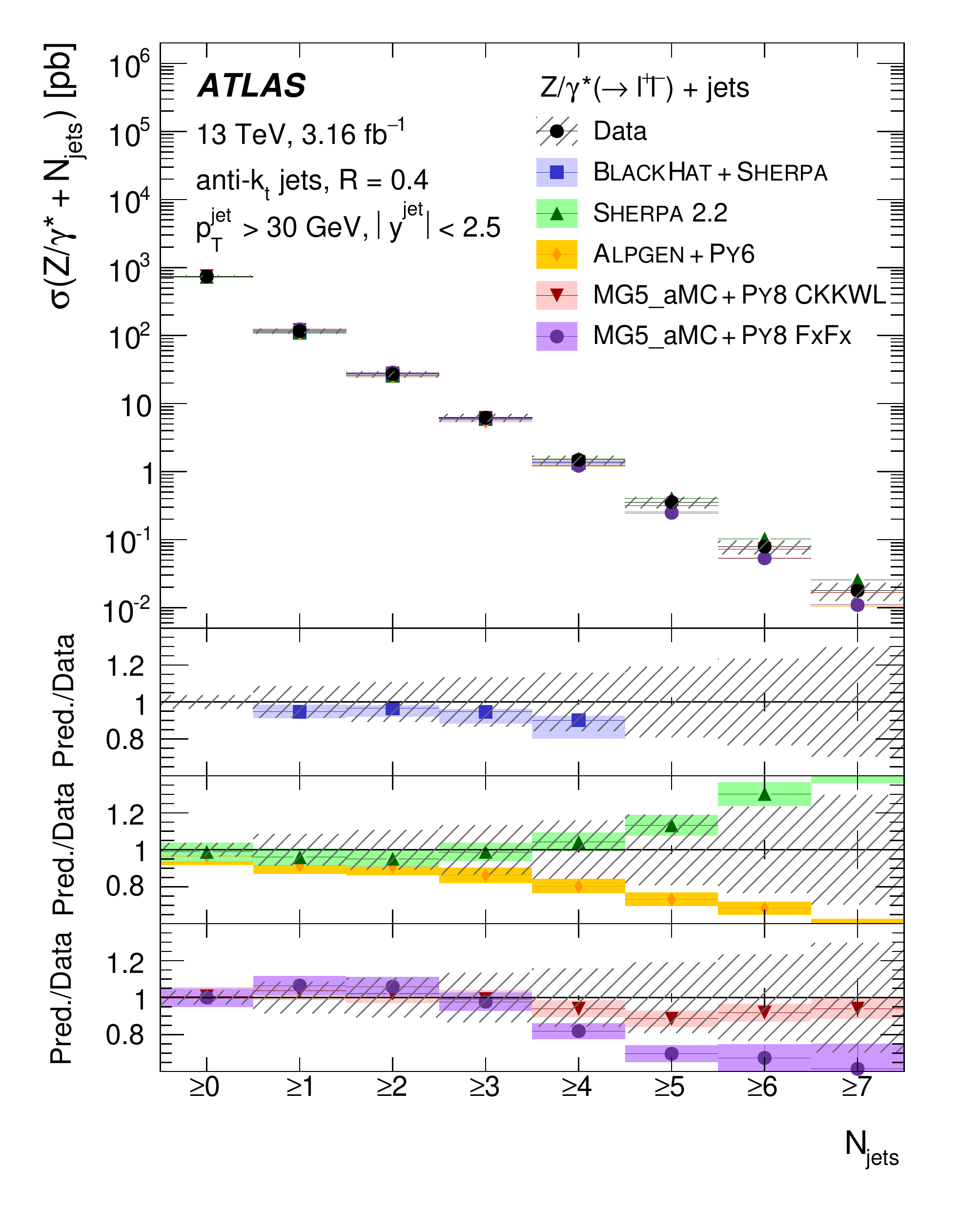}
\includegraphics[width=0.48\textwidth]{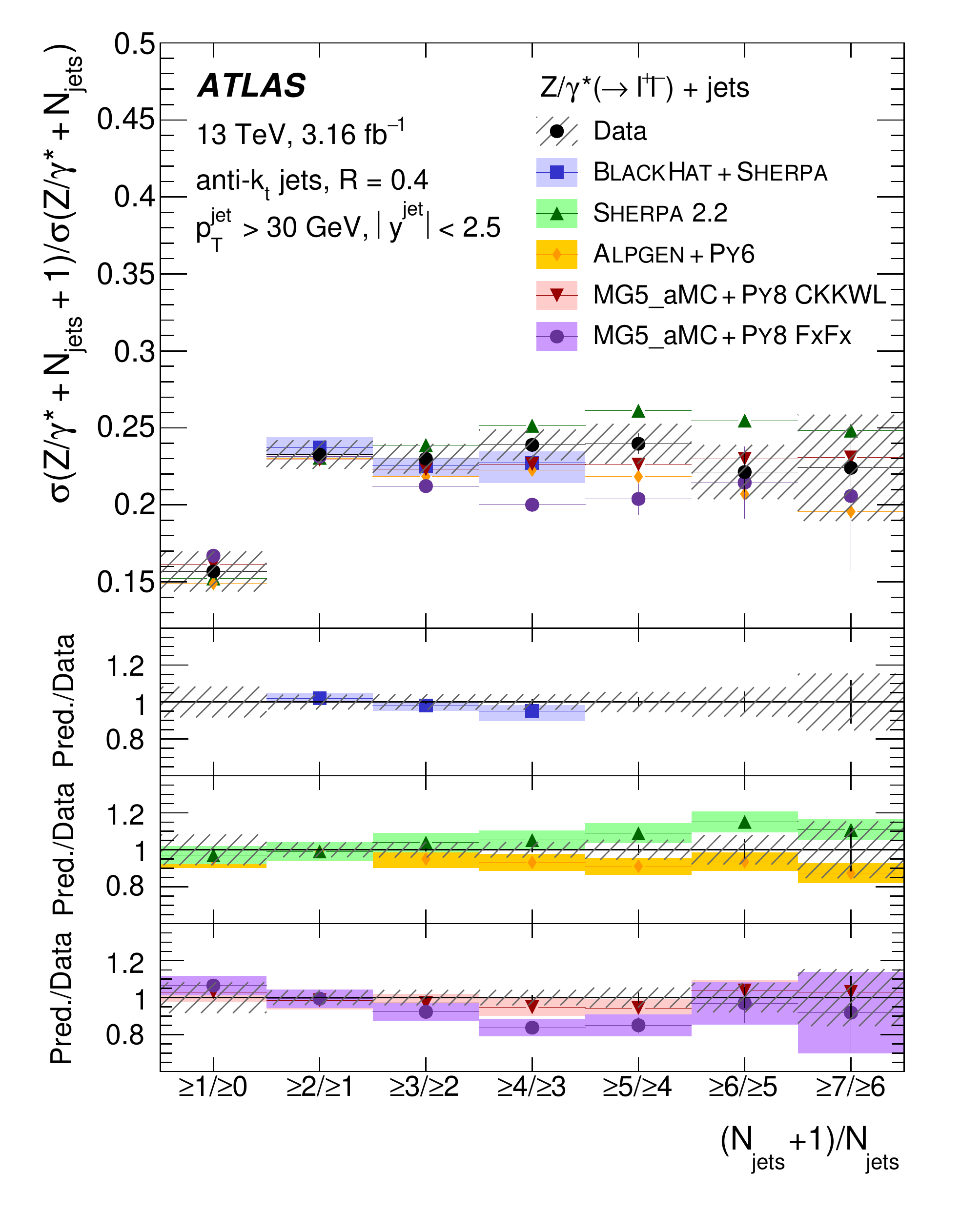}
\caption{Measured cross section as a function of the inclusive jet multiplicity (left) and ratio for successive  inclusive jet multiplicities (right) for inclusive \Zj events. 
The data are compared to the predictions from \BLACKHATSHERPA, \SHERPAtwotwo, \ALPGENPysix, \MGaMCPyCKKWL, and \MGaMCPyFxFx. The error bars correspond to the statistical uncertainty, and the hatched bands to the data statistical and systematic uncertainties (including luminosity) added in quadrature.
A constant  5\% theoretical uncertainty is used for \SHERPAtwotwo, \ALPGENPysix, \MGaMCPyCKKWL, and \MGaMCPyFxFx. Uncertainties from the parton distribution
 functions and QCD scale variations are included in the \BLACKHATSHERPA predictions, as described in Section~\ref{sec:dataMCFO}.}
\label{fig:xs_NinR_comb}
\end{center}
\end{figure} 
  
\subsection{Comparisons of results to predictions}
\label{sec:comparison}

The cross-section measurement for different inclusive \Zj multiplicities and their ratios obtained from the combination are found in Tables~\ref{tab:zjets_njet_comb} and~\ref{tab:zjets_njet_ratio_comb}. Figure~\ref{fig:xs_NinR_comb} shows the comparison of these results with the NLO QCD fixed-order calculations from \BLACKHATSHERPA and with the predictions from  \SHERPAtwotwo, \ALPGENPysix,  \MGaMCPyCKKWL, and \MGaMCPyFxFx. The plots show the particle-level cross section with the generator predictions normalised to the inclusive NNLO cross sections in the top panel, accompanied by the ratios of the various predictions with respect to the data in the bottom panels. Uncertainties from the parton distribution functions and QCD scale variations are included in the \BLACKHATSHERPA predictions, as described in Section~\ref{sec:dataMCFO}.
A constant  5\% theoretical uncertainty is used for \SHERPAtwotwo, \ALPGENPysix, \MGaMCPyCKKWL, and \MGaMCPyFxFx, as described in Table~\ref{tab:MC}.
The inclusive jet multiplicity decreases logarithmically while the ratio is flat in the presence of at least one jet.
The predictions are in agreement with the observed cross sections and their ratios, except for \SHERPAtwotwo, \ALPGENPysix and \MGaMCPyFxFx for high jet multiplicity, 
where a non-negligible fraction of the jets are produced by the parton shower.

\begin{table}[t]
\centering
\begin{tabular}{|c|r@{\,}@{$\pm$}@{\;}r@{\,}@{$\pm$}@{\;}r@{\,}@{$\pm$}@{\;}r|}
\hline\hline
Jet multiplicity & \multicolumn{4}{c|}{Measured cross section} \\
                 & \multicolumn{4}{c|}{$\pm$\,(stat.)\,$\pm$\,(syst.)\,$\pm$\,(lumi.) [pb]} \\
\hline\hline
& \multicolumn{4}{c|}{$Z \to \ell\ell$}\\
\hline\hline
$\geq\mathrm{0~jets}$ & \numRP{739.8599854}{0} & \numRP{0.6764300}{0} & \numRP{22.9845414}{0} & \numRP{15.6192920}{0} \\
$\geq\mathrm{1~jets}$ & \numRP{116.0000000}{1} & \numRP{0.2803500}{1} &  \numRP{9.6737257}{1} &  \numRP{2.4914574}{1} \\
$\geq\mathrm{2~jets}$ &  \numRP{26.9850006}{1} & \numRP{0.1057000}{1} &  \numRP{2.8482613}{1} &  \numRP{0.6013228}{1} \\
$\geq\mathrm{3~jets}$ &   \numRP{6.2037001}{2} & \numRP{0.0424930}{2} &  \numRP{0.8174381}{2} &  \numRP{0.1440070}{2} \\
$\geq\mathrm{4~jets}$ &   \numRP{1.4823000}{2} & \numRP{0.0191580}{2} &  \numRP{0.2309641}{2} &  \numRP{0.0357457}{2} \\
$\geq\mathrm{5~jets}$ &   \numRP{0.3554200}{2} & \numRP{0.0089666}{2} &  \numRP{0.0669050}{2} &  \numRP{0.0088749}{2} \\
$\geq\mathrm{6~jets}$ &   \numRP{0.0787690}{3} & \numRP{0.0040509}{3} &  \numRP{0.0178664}{3} &  \numRP{0.0020621}{3} \\
$\geq\mathrm{7~jets}$ &   \numRP{0.0177740}{4} & \numRP{0.0018656}{4} &  \numRP{0.0049071}{4} &  \numRP{0.0004867}{4} \\
\hline\hline
\end{tabular}
\caption{Measured combined fiducial cross sections for successive inclusive jet multiplicities.
The statistical, systematic, and luminosity uncertainties are given.}
\label{tab:zjets_njet_comb}
\end{table}

\begin{table}[t]
\centering
\begin{tabular}{|c|l@{\,}@{$\pm$}@{\;}l@{\,}@{$\pm$}@{\;}l@{\,}@{$\pm$}@{\;}l|}
\hline\hline
Jet multiplicity & \multicolumn{4}{c|}{Measured cross-section ratio} \\
                 & \multicolumn{4}{c|}{$\pm$\,(stat.)\,$\pm$\,(syst.)\,$\pm$\,(lumi.)} \\
\hline\hline
& \multicolumn{4}{c|}{$Z \to \ell\ell$}\\
\hline\hline
$\geq\mathrm{1~jets}~/\geq\mathrm{0~jets}$ & \numRP{0.1567600}{4} & \numRP{0.0004068}{4} & \numRP{0.0130991}{4} & \numRP{0.0000575}{4} \\
$\geq\mathrm{2~jets}~/\geq\mathrm{1~jets}$ & \numRP{0.2326700}{4} & \numRP{0.0010731}{4} & \numRP{0.0092790}{4} & \numRP{0.0001873}{4} \\
$\geq\mathrm{3~jets}~/\geq\mathrm{2~jets}$ & \numRP{0.2299200}{4} & \numRP{0.0018116}{4} & \numRP{0.0094696}{4} & \numRP{0.0002136}{4} \\
$\geq\mathrm{4~jets}~/\geq\mathrm{3~jets}$ & \numRP{0.2389500}{4} & \numRP{0.0034898}{4} & \numRP{0.0094214}{4} & \numRP{0.0002154}{4} \\
$\geq\mathrm{5~jets}~/\geq\mathrm{4~jets}$ & \numRP{0.2396800}{4} & \numRP{0.0067851}{4} & \numRP{0.0110975}{4} & \numRP{0.0002050}{4} \\
$\geq\mathrm{6~jets}~/\geq\mathrm{5~jets}$ & \numRP{0.2213400}{4} & \numRP{0.0126790}{4} & \numRP{0.0122535}{4} & \numRP{0.0002669}{4} \\
$\geq\mathrm{7~jets}~/\geq\mathrm{6~jets}$ & \numRP{0.2240300}{4} & \numRP{0.0264230}{4} & \numRP{0.0222400}{4} & \numRP{0.0002521}{4} \\
\hline\hline
\end{tabular}
\caption{Measured combined ratios of the fiducial cross sections for successive inclusive jet multiplicities. 
The statistical, systematic, and luminosity uncertainties are given.}
\label{tab:zjets_njet_ratio_comb}
\end{table}

\begin{figure}[t]
\begin{center}
\includegraphics[width=0.48\textwidth]{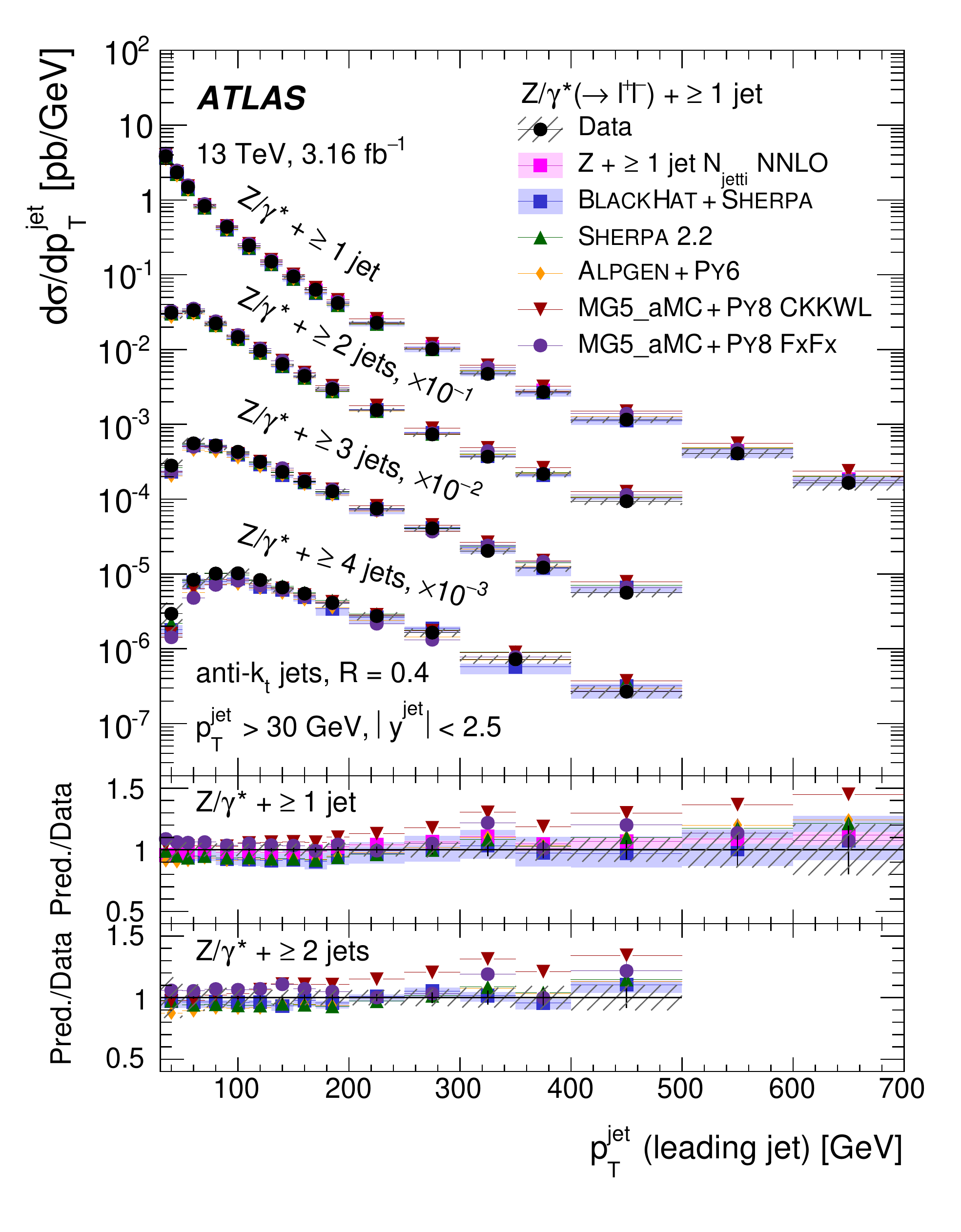}
\includegraphics[width=0.48\textwidth]{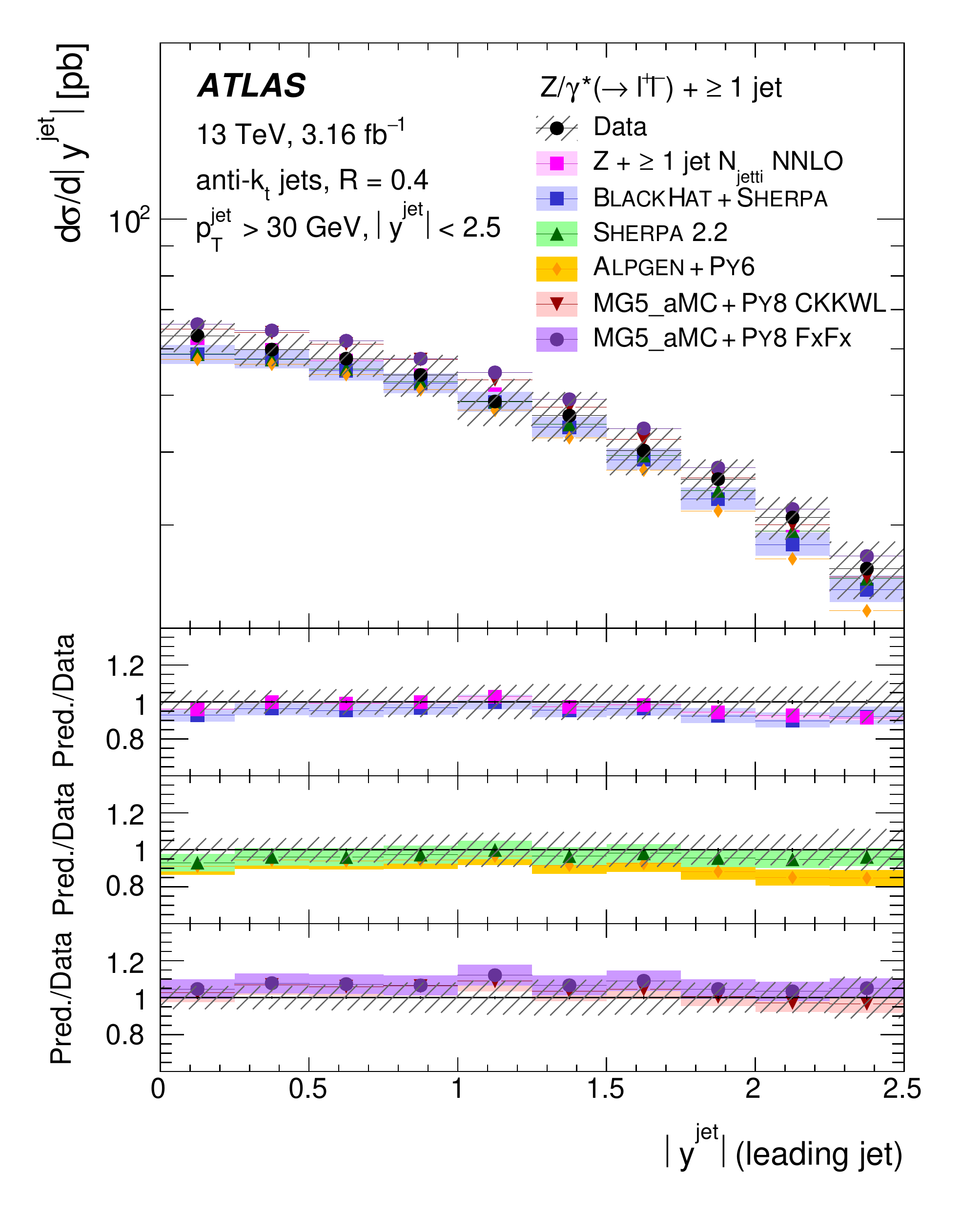}
\caption{Measured cross section as a function of the leading jet \pt\ for inclusive \Zonefourj events (left) and absolute value of the leading jet rapidity for inclusive \Zonej events (right). 
The data are compared to the predictions from \Njetti, \BLACKHATSHERPA, \SHERPAtwotwo, \ALPGENPysix, \MGaMCPyCKKWL, and \MGaMCPyFxFx. 
The error bars correspond to the statistical uncertainty, and the hatched bands to the data statistical and systematic uncertainties (including luminosity) added in quadrature. 
The details of the prediction uncertainties are given in the text.
For clarity, uncertainty bands are not shown for the Monte Carlo predictions in the left-hand plot.
Uncertainties from the QCD scale variations for the \Njetti predictions are included, as described in Section~\ref{sec:dataMCFO}. }
\label{fig:xs_Jet1PtNj_comb}
\end{center}
\end{figure}

\begin{figure}[t]
\begin{center}
\includegraphics[width=0.48\textwidth]{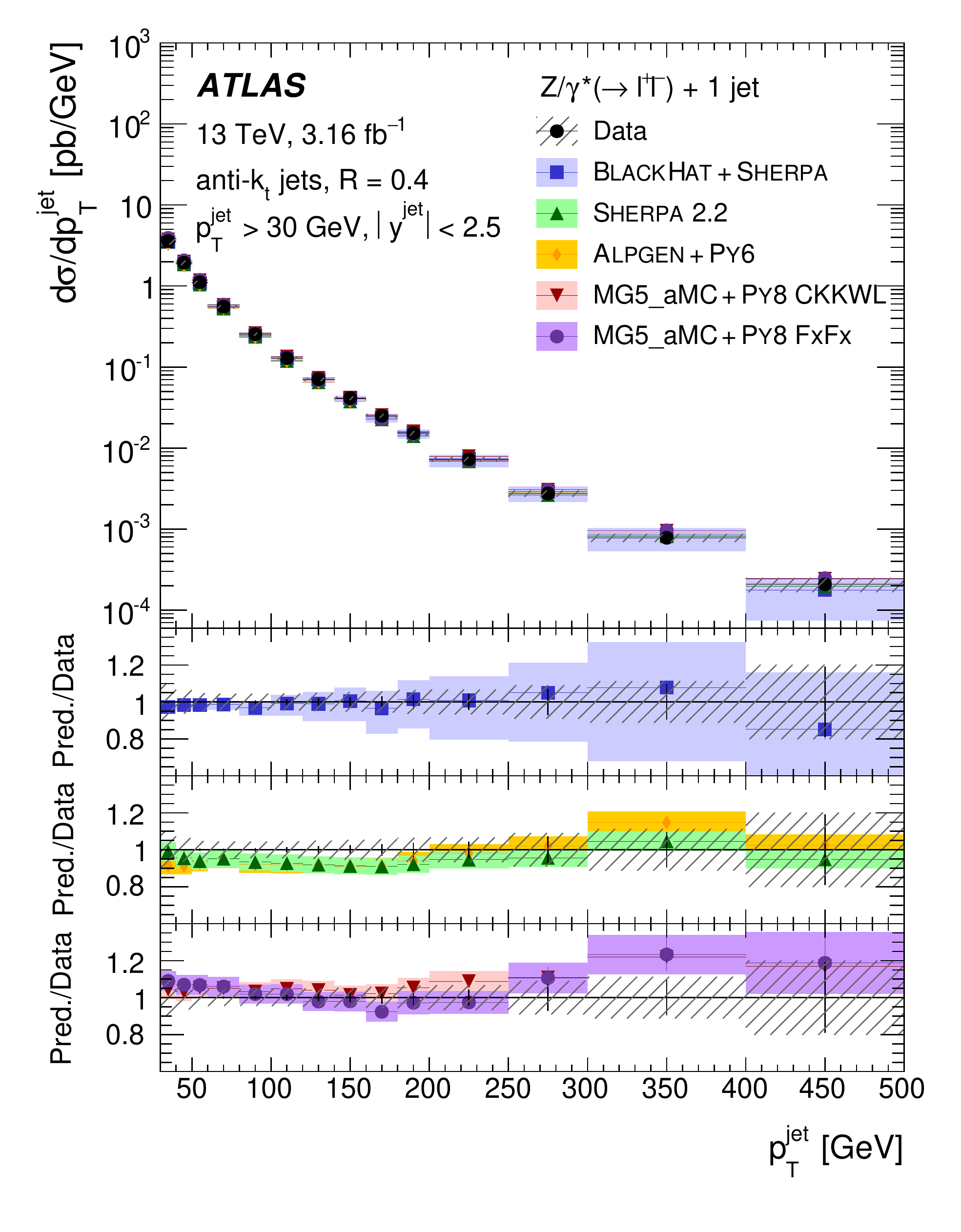}
\includegraphics[width=0.48\textwidth]{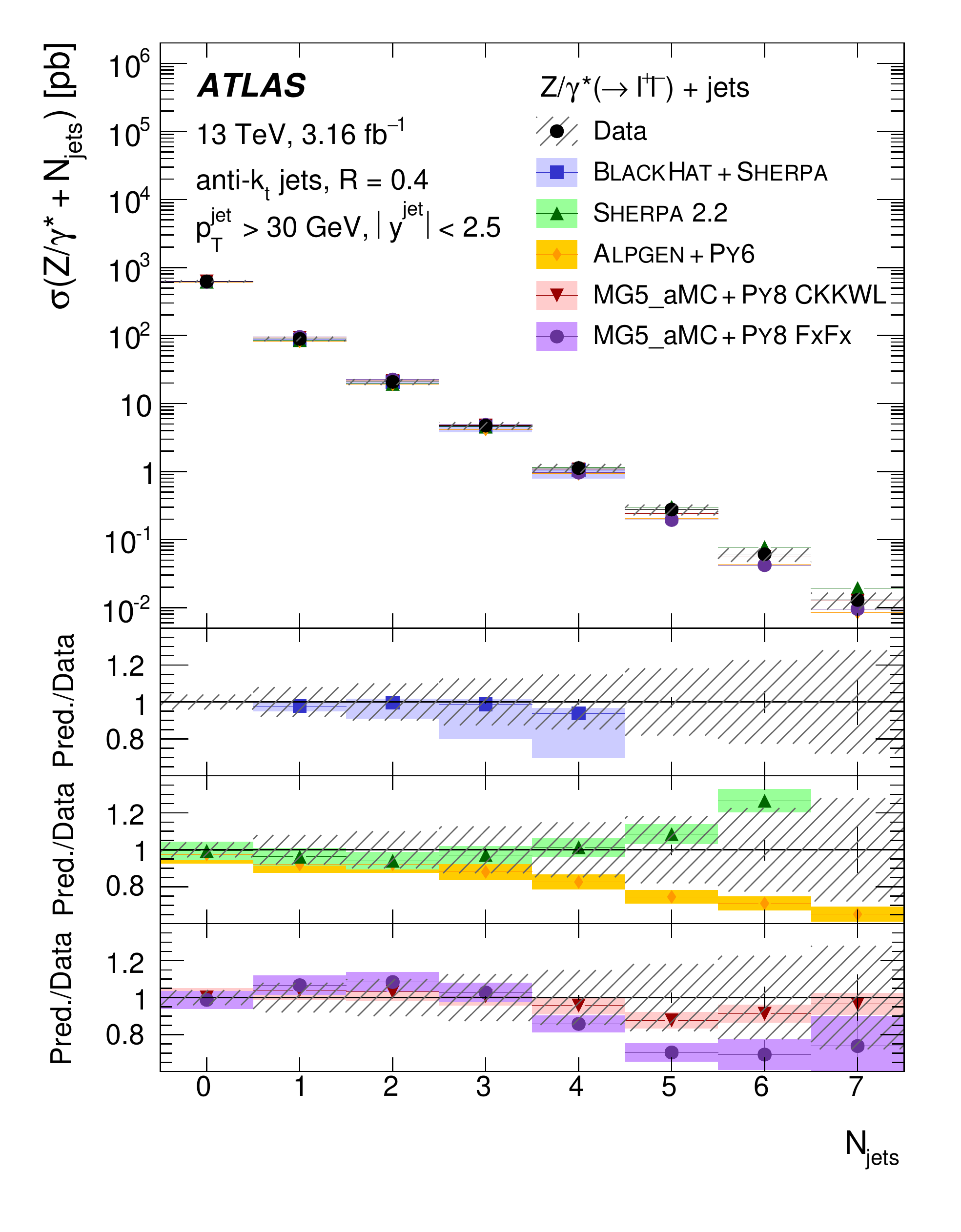}
\caption{Measured cross section as a function of jet \pt\ for exclusive \Zexclj events (left) and exclusive jet multiplicity (right). 
The data are compared to the predictions from \BLACKHATSHERPA, \SHERPAtwotwo, \ALPGENPysix, \MGaMCPyCKKWL, and \MGaMCPyFxFx. The error bars correspond to the statistical uncertainty, and the hatched bands to the data statistical and systematic uncertainties (including luminosity) added in quadrature. 
The details of the prediction uncertainties are given in the text.}
\label{fig:xs_Jet1PtNjexcl_comb}
\end{center}  
\end{figure}

The jet transverse momentum is a fundamental observable of the \Zj process and probes pQCD over a wide range of scales. Moreover, understanding the kinematics of jets in events with vector bosons associated with several jets is essential for the modelling of backgrounds for other SM processes and searches beyond the SM.
The leading jet \pt\ distribution (which is correlated with the \pT of the \Zboson boson) in inclusive \Zonefourj events is shown in Figure~\ref{fig:xs_Jet1PtNj_comb} and ranges up to 700~GeV. The LO generator \MGaMCPyCKKWL models a too-hard  jet \pt\ spectrum. This feature is known from studies of LO generators in $pp$ collisions at lower centre-of-mass energies~\cite{Zjets2013}, and can be interpreted as an indication that the dynamic factorisation and renormalisation scale used in the generation is not appropriate for the full jet \pt\ range.
In contrast, the predictions from  \BLACKHATSHERPA, \SHERPAtwotwo, and \MGaMCPyFxFx, which are based on NLO matrix elements,  are in agreement  with the measured cross section within the systematic uncertainties over the full leading jet \pt\ range. \ALPGENPysix also shows good agreement with the measured data. The \Njetti prediction models the spectrum for the \Zonej events well. Uncertainties from the QCD scale variations for the \Njetti predictions are included in the uncertainty band, as described in Section~\ref{sec:dataMCFO}. 
For the leading jet rapidity distribution in inclusive \Zonej events, also shown in this figure, all predictions show good agreement with the measured data within the uncertainties.

The exclusive jet \pt\ distribution probes the validity of  \Zexclj predictions at increasing QCD scales represented by the jet \pT\ in the presence of a jet veto at a constant low scale; for a jet \pt\ range of several hundred GeV, accessible with the current data set, the jet scale is of order ten times larger than the veto scale (30~GeV).
Figure~\ref{fig:xs_Jet1PtNjexcl_comb} demonstrates that all predictions studied are consistent with the data within systematic uncertainties over the full jet \pt\ range (up to 500~GeV).
This figure also shows the measured cross section as a function of the exclusive jet multiplicity, which decreases logarithmically. 
Similar trends as for the inclusive jet multiplicity (Figure~\ref{fig:xs_NinR_comb}) are observed.

\begin{figure}[t]
\begin{center}
\includegraphics[width=0.48\textwidth]{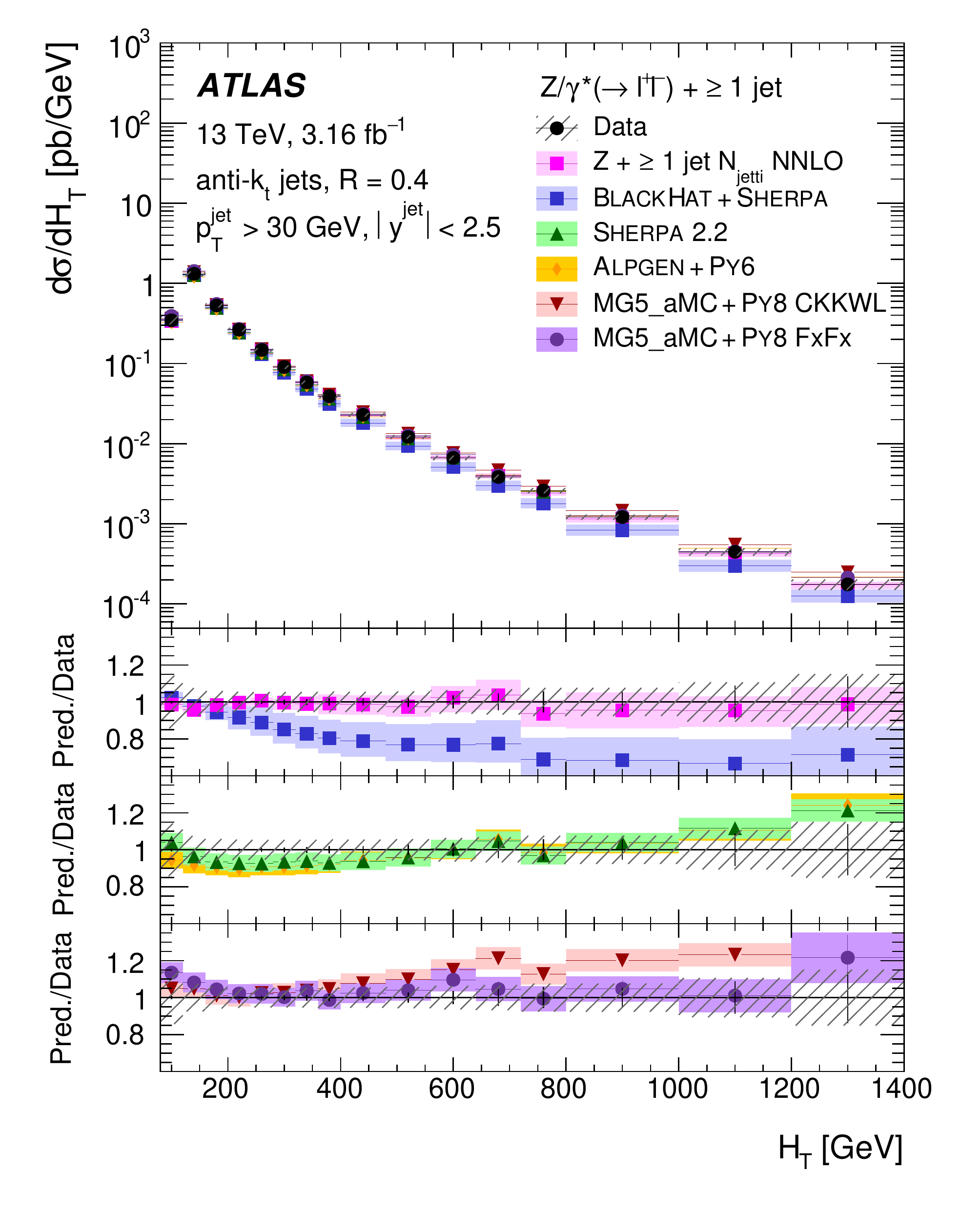}
\caption{Measured cross section as a function of \HT for inclusive \Zonej events. 
The data are compared to the predictions from  \Njetti, \BLACKHATSHERPA, \SHERPAtwotwo, \ALPGENPysix, \MGaMCPyCKKWL, and \MGaMCPyFxFx. The error bars correspond to the statistical uncertainty, and the hatched bands to the data statistical and systematic uncertainties (including luminosity) added in quadrature. The details of the prediction uncertainties are given in the text.}
\label{fig:xs_HT1jY_comb}
\end{center}
\end{figure}

Quantities based on inclusive \pT sums of final-state objects, such as \HT, the scalar \pT sum of all visible objects in the final state, are often employed in searches for physics beyond the Standard Model, to enrich final states resulting from the decay of heavy particles. The values \HT or  \HT/2 are also commonly used choices for scales for higher-order perturbative QCD calculations.  Large values for this quantity can result either from a small number of very energetic particles or from a large number of less energetic particles.  Figure~\ref{fig:xs_HT1jY_comb} shows the measured cross sections as a function of the \HT distribution (up to 1400~GeV) in inclusive \Zonej events. The predictions from \SHERPAtwotwo, \ALPGENPysix  and \MGaMCPyFxFx describe well the \HT distribution. The prediction from \MGaMCPyCKKWL describes well the turn-over in the softer part of the \HT spectrum, but overestimates the contribution at large values of \HT, in line with the overestimate of the cross sections for hard jets. The fixed-order \Zonej prediction from \BLACKHATSHERPA underestimates the cross section for values of $\HT>300$~GeV, as observed in similar measurements at lower centre-of-mass energies~\cite{Zjets2013,mySTDM-2012-24}, due to the missing contributions from events with higher parton multiplicities, which for large values of \HT constitute a substantial portion of the data. Agreement is recovered by adding higher orders in perturbative QCD, as demonstrated by the good description of \HT by \Njetti.

\begin{figure}[t]
\begin{center}
\includegraphics[width=0.48\textwidth]{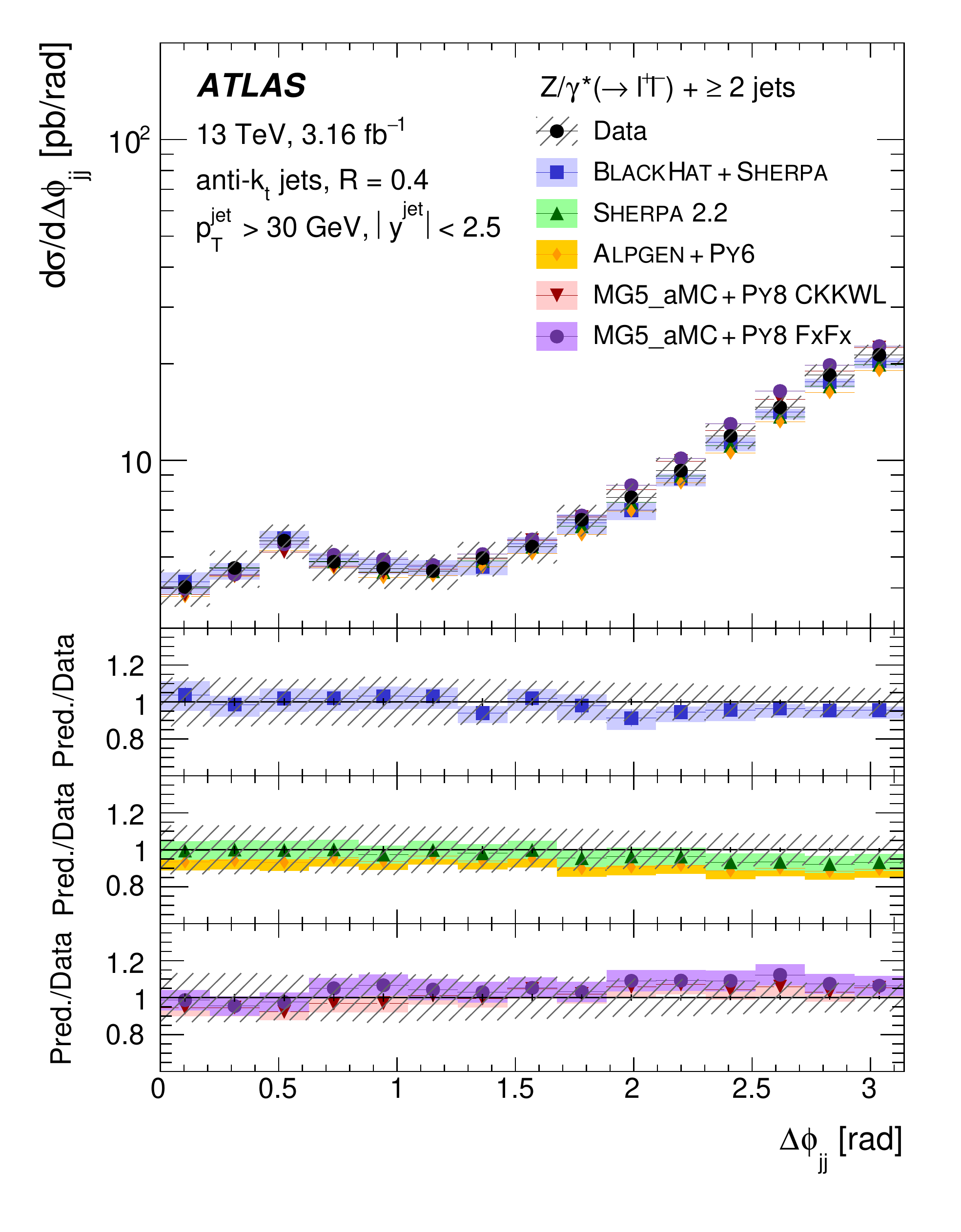}
\includegraphics[width=0.48\textwidth]{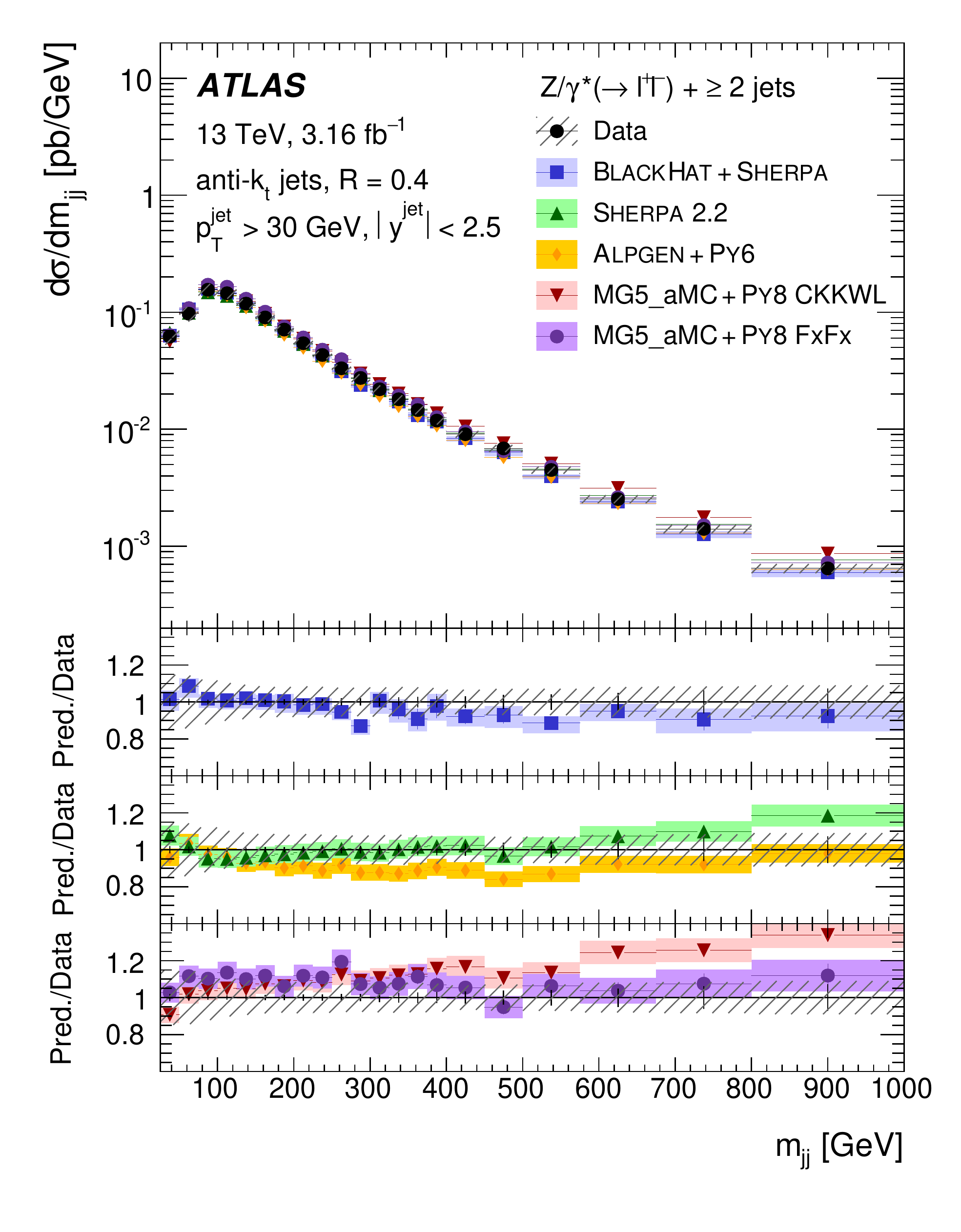}
\caption{Measured cross section as a function of \Dphijj (left) and \mjj (right) for inclusive \Ztwoj events. 
The data are compared to the predictions from \BLACKHATSHERPA, \SHERPAtwotwo, \ALPGENPysix, \MGaMCPyCKKWL, and \MGaMCPyFxFx. The error bars correspond to the statistical uncertainty, and the hatched bands to the data statistical and systematic uncertainties (including luminosity) added in quadrature. The details of the prediction uncertainties are given in the text.}
\label{fig:xs_JetDphiM_comb}
\end{center}
\end{figure}

Angular relations between the two leading jets and the dijet mass are frequently used to separate either heavier SM particles or beyond-SM physics from the \Zj process. 
Figure~\ref{fig:xs_JetDphiM_comb} shows the differential cross section as a function of azimuthal angular difference between the two leading jets for \Ztwoj events, \Dphijj. The tendency of the two jets to be back-to-back in the transverse plane is well modelled by all predictions. This figure also shows the measured cross sections as a function of the invariant mass \mjj of the two leading jets for \Ztwoj events. The shape of the dijet mass is modelled well by \BLACKHATSHERPA, \SHERPAtwotwo, \ALPGENPysix, and \MGaMCPyFxFx, whereas \MGaMCPyCKKWL shows a harder spectrum.

\clearpage

\section{Conclusion}
\label{sec:conclusion}

Proton--proton collision data at $\sqrt{s}=13$~TeV  from the LHC, corresponding to a total integrated luminosity of 3.16~fb$^{-1}$, have been analysed by the ATLAS collaboration to study events with \Zboson bosons decaying to electron or muon pairs, produced in association with one or more jets.
The fiducial production cross sections for \Zzerosevenj have been measured, within the acceptance region defined by $\pT^{\ell}>25$~GeV, $|\eta^{\ell}|<2.5$, $71<m_{\ell\ell}<111$~GeV, $\pTj > 30$~GeV, $|\yj|<2.5$, and $\Delta R(\ell,\text{jet}) > 0.4$, with a precision ranging from 4\% to 30\%. Ratios of cross sections for successive jet multiplicities and cross-section measurements as a function of different key variables such as the jet multiplicities, jet \pt\ for exclusive $Z+1$~jet events, leading jet \pt\ for \Zonefourj events, leading jet rapidity for \Zonej events, $H_{\text{T}}$, $\Delta\phi_{\text{jj}}$ and $m_{\text{jj}}$ have also been derived.

The measurements have  been compared to fixed-order calculations at NLO from \BLACKHATSHERPA and at NNLO from the \Njetti calculation, and to predictions from the generators 
\SHERPAtwotwo, \ALPGENPysix, \MGaMCPyCKKWL, and \MGaMCPyFxFx.
In general, the predictions are in good agreement with the observed cross sections and cross-section ratios within the uncertainties.
Distributions which are dominated by a single jet multiplicity are modelled well by fixed-order NLO calculations, even in the presence of a jet veto at a low scale.
The \textsc{ME+PS} generator \MGaMCPyCKKWL, which is based on LO matrix elements, models a too-hard jet spectrum, as observed in $\sqrt{s}=7$~TeV $pp$ collisions. It however models well the inclusive jet multiplicity distribution over the full multiplicity range. The modelling of the jet \pt and related observables is significantly improved by the \textsc{ME+PS@NLO} generators \SHERPAtwotwo  and \MGaMCPyFxFx, which use NLO matrix elements for up to two additional partons. The recent \Njetti predictions describe well key distributions such as the leading jet \pt\ and \HT.

The results presented in this paper provide essential input for the further optimisation of the Monte Carlo generators of \Zj production and constitute a powerful test of perturbative QCD for processes with a higher number of partons in the final state.

\section*{Acknowledgements}


We thank CERN for the very successful operation of the LHC, as well as the
support staff from our institutions without whom ATLAS could not be
operated efficiently.

We acknowledge the support of ANPCyT, Argentina; YerPhI, Armenia; ARC, Australia; BMWFW and FWF, Austria; ANAS, Azerbaijan; SSTC, Belarus; CNPq and FAPESP, Brazil; NSERC, NRC and CFI, Canada; CERN; CONICYT, Chile; CAS, MOST and NSFC, China; COLCIENCIAS, Colombia; MSMT CR, MPO CR and VSC CR, Czech Republic; DNRF and DNSRC, Denmark; IN2P3-CNRS, CEA-DSM/IRFU, France; SRNSF, Georgia; BMBF, HGF, and MPG, Germany; GSRT, Greece; RGC, Hong Kong SAR, China; ISF, I-CORE and Benoziyo Center, Israel; INFN, Italy; MEXT and JSPS, Japan; CNRST, Morocco; NWO, Netherlands; RCN, Norway; MNiSW and NCN, Poland; FCT, Portugal; MNE/IFA, Romania; MES of Russia and NRC KI, Russian Federation; JINR; MESTD, Serbia; MSSR, Slovakia; ARRS and MIZ\v{S}, Slovenia; DST/NRF, South Africa; MINECO, Spain; SRC and Wallenberg Foundation, Sweden; SERI, SNSF and Cantons of Bern and Geneva, Switzerland; MOST, Taiwan; TAEK, Turkey; STFC, United Kingdom; DOE and NSF, United States of America. In addition, individual groups and members have received support from BCKDF, the Canada Council, CANARIE, CRC, Compute Canada, FQRNT, and the Ontario Innovation Trust, Canada; EPLANET, ERC, ERDF, FP7, Horizon 2020 and Marie Sk{\l}odowska-Curie Actions, European Union; Investissements d'Avenir Labex and Idex, ANR, R{\'e}gion Auvergne and Fondation Partager le Savoir, France; DFG and AvH Foundation, Germany; Herakleitos, Thales and Aristeia programmes co-financed by EU-ESF and the Greek NSRF; BSF, GIF and Minerva, Israel; BRF, Norway; CERCA Programme Generalitat de Catalunya, Generalitat Valenciana, Spain; the Royal Society and Leverhulme Trust, United Kingdom.

The crucial computing support from all WLCG partners is acknowledged gratefully, in particular from CERN, the ATLAS Tier-1 facilities at TRIUMF (Canada), NDGF (Denmark, Norway, Sweden), CC-IN2P3 (France), KIT/GridKA (Germany), INFN-CNAF (Italy), NL-T1 (Netherlands), PIC (Spain), ASGC (Taiwan), RAL (UK) and BNL (USA), the Tier-2 facilities worldwide and large non-WLCG resource providers. Major contributors of computing resources are listed in Ref.~\cite{ATL-GEN-PUB-2016-002}.

\clearpage


\printbibliography

\newpage 
\begin{flushleft}
{\Large The ATLAS Collaboration}

\bigskip

M.~Aaboud$^\textrm{\scriptsize 137d}$,
G.~Aad$^\textrm{\scriptsize 88}$,
B.~Abbott$^\textrm{\scriptsize 115}$,
J.~Abdallah$^\textrm{\scriptsize 8}$,
O.~Abdinov$^\textrm{\scriptsize 12}$,
B.~Abeloos$^\textrm{\scriptsize 119}$,
R.~Aben$^\textrm{\scriptsize 109}$,
O.S.~AbouZeid$^\textrm{\scriptsize 139}$,
N.L.~Abraham$^\textrm{\scriptsize 151}$,
H.~Abramowicz$^\textrm{\scriptsize 155}$
% ...
\bigskip
\\
% ...
$^{8}$ Department of Physics, The University of Texas at Arlington, Arlington TX, United States of America\\
% ...
$^{12}$ Institute of Physics, Azerbaijan Academy of Sciences, Baku, Azerbaijan\\
% ...
$^{88}$ CPPM, Aix-Marseille Universit{\'e} and CNRS/IN2P3, Marseille, France\\
% ...
$^{109}$ Nikhef National Institute for Subatomic Physics and University of Amsterdam, Amsterdam, Netherlands\\
% ...
$^{115}$ Homer L. Dodge Department of Physics and Astronomy, University of Oklahoma, Norman OK, United States of America\\
% ...
$^{119}$ LAL, Univ. Paris-Sud, CNRS/IN2P3, Universit{\'e} Paris-Saclay, Orsay, France\\
% ...
$^{137}$ $^{(a)}$ Facult{\'e} des Sciences Ain Chock, R{\'e}seau Universitaire de Physique des Hautes Energies - Universit{\'e} Hassan II, Casablanca; $^{(b)}$ Centre National de l'Energie des Sciences Techniques Nucleaires, Rabat; $^{(c)}$ Facult{\'e} des Sciences Semlalia, Universit{\'e} Cadi Ayyad, LPHEA-Marrakech; $^{(d)}$ Facult{\'e} des Sciences, Universit{\'e} Mohamed Premier and LPTPM, Oujda; $^{(e)}$ Facult{\'e} des sciences, Universit{\'e} Mohammed V, Rabat, Morocco\\
% ...
$^{139}$ Santa Cruz Institute for Particle Physics, University of California Santa Cruz, Santa Cruz CA, United States of America\\
% ...
$^{151}$ Department of Physics and Astronomy, University of Sussex, Brighton, United Kingdom\\
% ...
$^{155}$ Raymond and Beverly Sackler School of Physics and Astronomy, Tel Aviv University, Tel Aviv, Israel\\
% ...
$^{a}$ Also at Department of Physics, King's College London, London, United Kingdom\\
$^{b}$ Also at Institute of Physics, Azerbaijan Academy of Sciences, Baku, Azerbaijan\\
$^{c}$ Also at Novosibirsk State University, Novosibirsk, Russia\\
$^{d}$ Also at TRIUMF, Vancouver BC, Canada\\
$^{e}$ Also at Department of Physics {\&} Astronomy, University of Louisville, Louisville, KY, United States of America\\
% ...
\end{flushleft}


\end{document}